\documentclass[graybox]{svmult}

\usepackage{mathptmx}       
\usepackage{helvet}         
\usepackage{courier}        
\usepackage{type1cm}        
\usepackage{amssymb,physics}
\usepackage{makeidx}         
\usepackage{graphicx}        
\usepackage{multicol}        
\usepackage[bottom]{footmisc}
\usepackage{hyperref}        
\usepackage{soul}            
\hypersetup{colorlinks=true,urlcolor=blue}
\usepackage{aas_macros}
\usepackage[square,numbers]{natbib}

\usepackage{xcolor}

\newcommand{\hbindex}[1]{\hl{#1}\index{#1}}  
\makeindex             

\begin{document}

\def\lum {\mbox{erg\,s$^{-1}$}}
\def\flux {\mbox{erg\,s$^{-1}$\,cm$^{-2}$}}
\def\kt {\mbox{$kT$}}

\def\AB#1{{\textcolor{orange}{#1}}}
\def\PE#1{{\textcolor{blue}{#1}}}

\title*{Isolated Neutron Stars}
\author{Alice Borghese\thanks{corresponding author} and Paolo Esposito}

\institute{Alice Borghese \at Instituto de Astrofísica de Canarias, E-38205 La Laguna, Tenerife, Spain; Departamento de Astrofísica, Universidad de La Laguna, E-38206 La Laguna, Tenerife, Spain; Institute of Space Sciences (ICE, CSIC), Campus UAB, Carrer de Can Magrans s/n, 08193, Barcelona, Spain;
\email{alice.borghes@gmail.com}
\and Paolo Esposito \at Scuola Universitaria Superiore IUSS Pavia, Palazzo del Broletto, piazza della Vittoria 15, 27100 Pavia, Italy; INAF--Istituto di Astrofisica Spaziale e Fisica Cosmica di Milano, via A.\,Corti 12, 20133 Milano, Italy; \email{paolo.esposito@iusspavia.it}}
%
%
\maketitle

\abstract{Non-accreting neutron stars display diverse characteristics, leading us to classify them into several groups. This chapter is an observational driven review in which we survey the properties of the different classes of isolated neutron stars: from the `normal'rotation-powered pulsars, to magnetars, the most magnetic neutron stars in the Universe we know of; from central compact objects (sometimes called also anti-magnetars) in supernova remnants, to the X-ray dim isolated neutron stars. We also highlight a few sources that have exhibited features straddling those of different sub-groups, blurring the apparent diversity of the neutron star zoo and pointing to a gran unification.
}

\section*{Keywords} 
Isolated Neutron Stars; Rotation-powered Pulsars; Magnetars; X-ray Dim Isolated Neutron Stars; Central Compact Objects.

\section{Introduction}
\label{sec:intro}

\begin{figure}
\centering
\includegraphics[width=1\columnwidth]{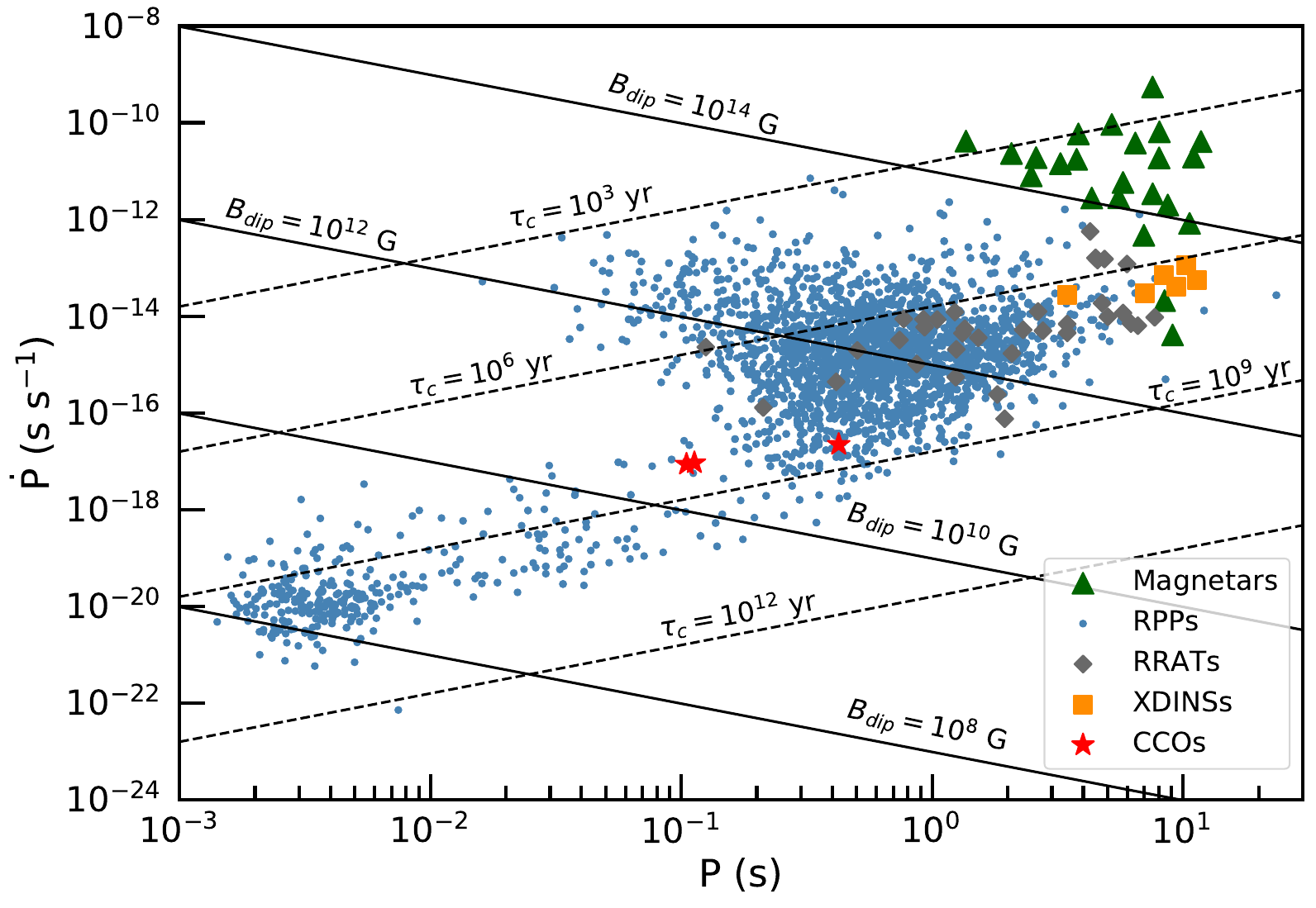} 
\caption{\label{fig:p_pdot} $P$--$\dot{P}$ diagram with the non-accreting neutron stars included in the ATNF pulsar catalogue v.1.66. Green triangles, blue dots, grey diamonds, orange squares and red stars represent magnetars, rotation-powered pulsars, RRATs, X-ray dim isolated neutron stars and central compact objects, respectively. Also plotted are lines of constant dipolar magnetic field (solid lines) and characteristic age (dashed lines) as derived from equations \ref{eq:Bdip} and \ref{eq:tau}. Image credit: A.\,Borghese.
} 
\end{figure}

After Chandrasekhar \cite{Chandrasekhar1931} argued the existence of a maximum mass for white dwarfs above which they would collapse gravitationally in 1930, the discovery of the neutron by Chadwick \cite{Chadwick1932} led astrophysics to conceive the existence of an even more extreme endpoint of the stellar evolution: a neutron star. Baade and Zwicky \cite{Baade1934} were the first to observe that supernova explosions could be the violent  transition of normal stars to neutron stars. 

The earliest electromagnetic signal from a neutron star was detected in X-rays, from the first X-ray source identified outside the Solar System, Scorpius\,X-1, in a legendary experiment led by Giacconi \cite{Giacconi1962}. Shklovsky \cite{Shklovsky1967} correctly interpreted it as a neutron star in the process of accreting matter. However, neutron stars become a widely-accepted astrophysical reality only with the discovery of the \emph{pulsating radio sources}, in short \hbindex{\emph{pulsars}} \cite{hewish68} (pulsating, because in the discovery paper radial pulsations from a compact star, either a white dwarf or a neutron star, were suggested as the most plausible physical phenomenon to explain signals recurring on the scale of seconds). The subsequent measure of the slow-down of the rate of pulses from the Crab pulsar \cite{Richards1969} firmly identified pulsars with neutron stars. Incidentally, the timing parameters of the Crab pulsar also showed that the modulation of the radio signal was actually due to the neutron star rotation rather than its oscillations/pulsations. The periodic modulation arises when the emission pattern is asymmetric or its axis of symmetry is misaligned with the rotation axis.

The term pulsar, which first appeared in the popular press in 1968 \cite{Jaeger2023}, was a portmanteau of ``pulsating quasar''. It is clearly non a good description of these sources, but it sticked. Also, the modulation of the emission from these objects has now been detected in all wavelengths, from radio to gamma-rays, and even radio-quiet `pulsars' were discovered in X-rays and gamma-rays!
The population of these celestial objects now totals more than 3000 sources, either isolated or in a binary system.\footnote{We shall also call `isolated' or 'solitary' any neutron star that is not accreting matter from the companion star, regardless of it being truly isolated or in a binary/multiple system. We will not discuss accreting neutron stars, which are covered in other chapters.} The number is constantly increasing thanks to intensive radio surveys with unprecedented sensitivity, all-sky X-/$\gamma$-ray monitors, and multi-wavelength observations targeted at supernova remnant and pulsar wind nebulae candidates.

Neutron stars display diverse observational properties broadly according to the primary source of energy that feeds their emission. These are:
\begin{itemize}
\item {\it Rotation}: Rotation-powered pulsars (RPPs) are driven by the loss of kinetic rotational energy due to the braking caused by their magnetic field. Most of the known radio pulsars are RPPs, and they are associated with nebular structures arising from a relativistic wind of particles emitted by the neutron star. A RPP can be found isolated or in a binary system.

\item {\it Cooling}: Thermally-powered neutron stars store their energy as internal heat, which can either be the leftover from the neutron star formation (only observable for the first $\sim$10$^5$\,yr) or the results of surface reheating from external sources, such as returning currents in the magnetosphere. Moreover, strong magnetic fields can provide a source of heat via Joule dissipation of the electrical currents that circulate in the neutron star crust, thus powering thermal emission.

\item {\it Magnetic energy}: Magnetic powered neutron stars generally possess X-ray luminosities larger than their rotational energy loss rates. The decay and instabilities of their strong magnetic fields (up to 10$^{15}$\,G at the surface) is believed to be the engine of the persistent emission and bursting activities of these sources. The magnetic field evolution causes magnetic stresses, triggering instabilities which give rise to transient phenomena, such as outbursts or short bursts (see Sec.\,\ref{subsec:transient_em}). These sources used to be classified as either `anomalous X-ray pulsars' or `soft gamma repeaters', but now they are often referred to as magnetars, from the model proposed in the 1990s.

\end{itemize}

These different emission mechanisms lead to different neutron star classes, such as the already mentioned RPPs and magnetars, or the thermally-emitting X-ray dim isolated neutrons stars (XDINSs). For the central compact objects (CCOs), it is less clear, but we note that the three energy sources are not mutually exclusive. Observational evidence that the boundaries between the groups are more blurred than we previously thought is mounting, prompting the possibility of a {\it grand unification} of the observational manifestations of isolated neutron stars.

As mentioned above, the loss of rotational kinetic energy is one of the primary source of energy for rotating neutron stars. Isolated rotation-powered neutron stars in the Galaxy are all observed to spin down, releasing energy at a rate $ \abs{\dot{E}} = I \omega \abs{\dot{\omega}} = 4 \pi^{2} I P^{-3} \dot{P}$, where $P$ is the spin period, $\dot{P}$ is its time derivative, $I$ is the moment of inertia and $\omega=2\pi P^{-1}$ is the angular frequency. Therefore, a measurement of $P$ and $\dot{P}$ determines the energy budget available to power the emission, the so-called spin down luminosity:
\begin{equation}
    \abs{\dot{E}} = I \omega \abs{\dot{\omega}} = 4 \pi^{2} I P^{-3} \dot{P} \simeq 4 \times 10^{33} P^{-3} \dot{P}_{13}\;\rm{erg\;s^{-1}},  
    \label{eq:Edot}
\end{equation} 
where $I=10^{45}$\,g\,cm$^{2}$ assuming standard parameter for neutron stars (mass $M=1.4M_\odot$ and radius $R=10$\,km), $P$ is expressed in seconds, and $\dot{P}_{13}$ is the spin derivative in units of 10$^{-13}$\,s\,s$^{-1}$.

According to the {\it magnetic dipole model} (see e.g., \citep{Pacini1967}), a neutron star rotates in vacuum at an angular frequency $\omega$ and possesses a magnetic dipole momentum $\mu$ oriented at an angle $\alpha$ with the rotation axis. The intensity of the magnetic dipole can be parametrized in terms of the dipole magnetic field at the equator $ B_{\rm dip}$, $\abs{\mu} = B_{\rm dip} R^3$. Given the misalignment between the magnetic and rotational axes, such rotating magnetic dipole radiates energy at a rate given by the Larmor formula   
\begin{equation}
    \dot{E}_\mathrm{dip}= - \frac{2}{3c^3} \abs{\ddot{\mu}}^2 = - \frac{2}{3c^3} \mu^2 \omega^4 (\sin{\alpha})^2 =
    - \frac{2}{3c^3} B^2_\mathrm{dip} R^6 \omega^4 (\sin{\alpha})^2.
    \label{eq:Edotdip}
\end{equation}

If, for simplicity, one considers that the rotational and magnetic axes are orthogonal ($\alpha=90^\circ$) and that magnetic dipole radiation is the dominant physical mechanism behind energy losses, the strength of the dipolar magnetic field at the magnetic equator can be estimated by equating Equation\,\ref{eq:Edot} with Equation\,\ref{eq:Edotdip}:

\begin{equation}
    B_{\rm dip}  \approx \bigg( \frac{3 c^3}{8\pi^2} \frac{I}{ R^6}\bigg)^{1/2} (\dot{P}P)^{1/2} \simeq 3.2 \times 10^{19} (P \dot{P})^{1/2}\; \rm{G}. 
    \label{eq:Bdip}
\end{equation}
For any power-law deceleration model, such as the magnetic dipole model, a general spin down formula of the form $\dot{\nu} \propto \nu^{n} $ can be written, where $\nu=P^{-1}$ is the spin frequency and $n \propto \nu \ddot{\nu} \dot{\nu}^{-2}$ is the braking index, determined by the torque mechanism working against the star rotation (e.g., $n = 3$ for pure magnetic dipole radiation, $n = 5$ for dominant gravitational radiation\footnote{A neutron star can spin down through the emission of gravitational waves if it possesses a time-varying quadrupole moment.}). By integrating the spin down formula, we obtain the neutron star age
\begin{equation}
    \tau = \frac{P}{(n-1) \dot{P}} \left[ 1 - \left( \frac{P_0}{P} \right)^{n-1} \right],
\end{equation}
where $P_0$ is the spin period at birth. Under the assumption that the spin period at birth is much shorter than the current value, and that the spin-down is due to the magnetic dipole radiation ($n = 3$), $\tau$ simplifies to the so-called characteristic age
\begin{equation}
    \tau_{\rm c} = \frac{P}{2 \dot{P}}.
    \label{eq:tau}
\end{equation}
Note that $\tau_{\rm c}$ does not necessarily provide a reliable estimate of the true age of the neutron star because of the many assumptions and simplifications, but for most pulsars it is the only available appraisal.

Although the magnetic-dipole braking model assumes that the star rotates in vacuum, which is not the case, it yields estimates of important physical parameters with a simple measurement of $P$ and $\dot{P}$. For this reason, the $P$--$\dot{P}$ diagram is a useful tool. Fig. \ref{fig:p_pdot} shows the $P$--$\dot{P}$ phase space with the non-accreting neutron stars included in the Australia Telescope National Facility (ATNF) Pulsar Catalogue, version 1.66\footnote{\url{https://www.atnf.csiro.au/research/pulsar/psrcat/index.html}, \cite{Manchester05}.}. Lines of constant $B_{\rm dip}$ and $\tau_{\rm c}$, as derived from Equations\,\ref{eq:Bdip} and \ref{eq:tau} respectively, are also indicated. The RPPs (blue dots) represent the bulk of the \emph{known} neutron star population, with typical dipolar magnetic fields $B_{\rm dip} \sim 10^{12}$\,G and characteristic ages $\tau_{\rm c} \sim 10^{7}$\,yr. The CCOs (red stars) blend with the radio pulsars, but their dipolar fields are lower ($B_{\rm dip} \sim 10^{10}$\,G). In the top right corner, there are two groups of neutrons stars, the magnetars (green triangles) and the XDINSs (orange squares), with $\tau_{\rm c} \sim 10^3 - 10^6$\,yr and $B_{\rm dip} \sim 10^{13} - 10^{15}$\,G. In the following, we will discuss in some detail the observational features of these isolated neutron star classes.

\section{Rotation-Powered Pulsars}
\label{sec:psr}

\hbindex{Rotation-Powered Pulsars} (RPPs) are usually discovered at radio frequencies, although they typically channel only a negligible fraction of their spin-down energy loss in that energy band ($\approx10^{-7}$--$10^{-5}$, and a similar fraction is observed in optical).
Indeed, way more power is channelled in the X-ray band ($\approx10^{-4}$--$10^{-3}$), and even a larger amount in gamma-rays ($\approx10^{-2}$--$10^{-1}$).\footnote{Furthermore, the observed braking index values are in general smaller than 3, the figure expected for pure dipolar braking, suggesting that a fraction of the spin-down energy loss is carried away by the pulsar wind, which may result in nebular emission.} Soft X-rays, however, are optimally suited to probe the thermal emission of neutron stars. Decades of observations have most certainly showed that thermal as well as non-thermal mechanisms of emission are at work in pulsars. Their relative importance is dependent on the age of the RPPs, and also on its magnetic field.

The most obvious source of thermal radiation for a neutron star is the residual heat from its formation in the supernova explosion, at $T>10^{10}$\,K (when the iron-group nuclei in the core separate in $\alpha$ particles, free nucleons, etc.). In the course of hours, much of the heat of the proto-neutron star is radiated away by neutrinos emitted through a variety of processes, while the temperature decreases to $\approx10^{9}$--$10^{10}$\,K. After $\approx$10--100\,yr (the time-scale of the thermal relaxation), the internal layers are expected to have become isothermal. At this point, the thermal evolution of the neutron star essentially depends on its heat capacity and the balance between the energy losses (mainly due to the emission of neutrinos from the interior and of photons from the surface) and heating mechanisms, such as the decay of the magnetic field, frictional heating, and (exothermal) nuclear reactions. The cooling is dominated by neutrino emission until the central temperature decreases to $\approx10^{5}$--$10^{7}$\,K, which may take $\approx10^{3}$--$10^{6}$\,yr. While most cooling curves agree that neutron stars reach a temperature of $\approx10^{5}$\,K in $10^{7}$\,yr, the path changes dramatically between the \emph{minimal} (or \emph{standard}) cooling, in which the temperature decreases gradually (via modified Urca reactions), and \emph{enhanced} (or \emph{accelerated} cooling), where direct Urca processes still take place (owing to very high central density and/or exotic composition). In the latter scenario, the temperature drops abruptly by a factor $\approx$10 during the first $\approx$50--100\,yr. In general, a large number of different cooling curves can be found, as the thermal evolution of the neutron stars critically depends on a number of poorly known physical conditions, including the neutron star mass, composition, equation of state, the impact of superfluidity and superconductivity, and the magnetic field on the various processes, as well as the processes themselves (a comprehensive overview of the neutron star cooling is given in \cite{Potekhin2015}). Therefore, the importance of constraining models with measurements of the thermal component is clear. The thermal component can be observed in high-quality spectra from the extreme ultraviolet to soft X-rays (roughly below 0.5\,keV).

Naively, thermal emission can be expected to be isotropic. For RPPs, however, the situation is more complicated. For instance, heating of polar areas by the relativistic particles accelerated in strong magnetic field, influence of the magnetic field on thermal conductivity of the crust and/or propagation of energy through magnetised atmosphere might lead to apparently non-uniform surface temperature distribution. As a result, isolated neutron stars are often observed to exhibit pulsed X-ray emission, typically with sinusoidal profiles with small pulsed fraction. In general, the presence of an atmosphere (0.1--10\,cm thick, with density $\lesssim10^2$\,g\,cm$^{-3}$, gaseous or condensed, depending on the chemical composition, temperature and magnetic field strength\footnote{In general, low temperatures and high magnetic fields favour a condensed atmosphere.}) is expected to significantly alter the emerging spectrum (see \cite{Potekhin2015DLP} for a review). Heavy-element atmospheres should result in the presence of many spectral features, due to cyclotron and absorption lines by ions in different ionization states. However, if observed at low spectral resolutions, such a spectrum appears similar to a blackbody of comparable effective temperature.  
Light elements accreted from the interstellar medium or from the fall-back of supernova debris have a significant impact on the spectral continuum, resulting in a high-energy tail and  producing few absorption features (only cyclotron lines in the case of complete ionization). It has been estimated that even a tiny amount of hydrogen ($\gtrsim10^{-20}$\,M$_\odot$) makes the emerging spectrum virtually identical to that of a pure hydrogen atmosphere.
Essentially, this is due to the fact that the opacity of light elements decreases rapidly with the energy, so that at high energies only hotter deeper layers of the neutron star can be observed. Because of this, the spectral parameters derived by a fit with a simple blackbody model can be wrong: the temperature can be overestimated by a factor up to $\approx$3 and the blackbody radius can be substantially overvalued (a number of different neutron star atmosphere models are available in the main fitting packages for X-ray spectral analysis). 
The magnetic field substantially changes the opacities for the different polarizations of the photons, thus making the emerging flux dependent on the direction of the magnetic field, and producing a flux modulation with the spin in non-aligned rotating neutron stars even in the unlikely case of perfectly uniform surface temperature.

Thermal emission may also result from the presence of hot spots on the neutron star surface due to currents of relativistic pairs of electrons and positrons from the materialization of gamma-rays produced in the polar-cap or outer-gap acceleration sites in the magnetosphere. This thermal emission is generally rather hard (0.1--0.2\,keV) and traceable to a blackbody radius much smaller than the stellar radius (from just a few meters to 1--2\,km). It is often the dominating thermal component, especially in old objects, such as the recycled millisecond pulsars,\footnote{Recycled millisecond pulsars are very old (they are particularly frequent in globular clusters) and rapidly-spinning pulsars ($P<0.1$\,s) with weak magnetic fields ($B\approx10^{10}$--$10^{11}$\,G). The main (and well-established) scenario to explain their paradoxical combination of short period and old age states that they are the descendants of low-mass neutron star accreting binary systems in which the neutron star, after a Gyr-long phase of mass (and angular momentum) transfer via a disc, starts to shine as a RPP when the mass transfer rate decreases or ceases. The reason of their low magnetic field is less certain but possible answers are the decay of the field with the age and/or that the accreted matter screened and buried the crustal field.} which have already lost much of their initial heat. Unless the system geometry and the viewing angle are unfavourable, these hot spots obviously produce a flux modulation, and thus their emission can be disentangled from the other components via phase-resolved spectroscopic studies (e.g. \cite{DeLuca2005}). \\

\begin{figure}
\centering
\includegraphics[width=0.7\columnwidth]{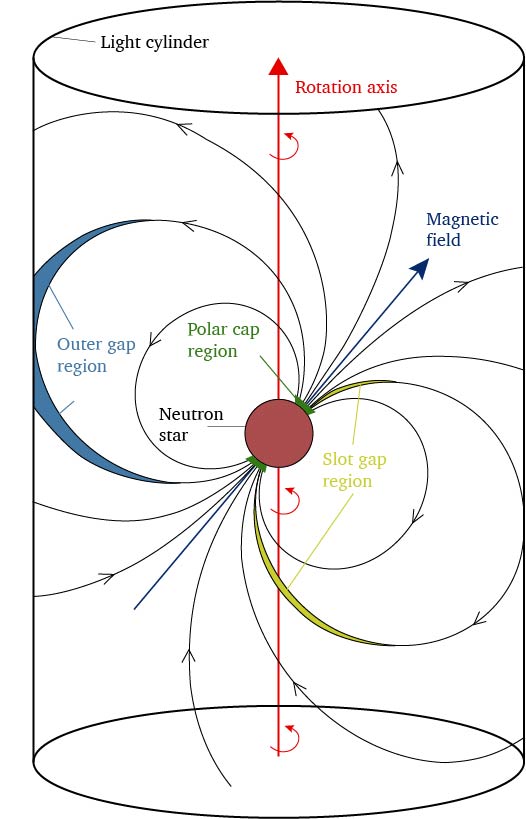} 
\caption{Scheme of the magnetosphere of a pulsar.\label{fig:crab_magnetosphere}}
\end{figure}

Pulsar electrodynamics is an immense topic and we refer the readers to \cite{Harding2021} and references therein for an overview. As anticipated in the introduction, there is a general agreement on the fact that rotating and magnetized neutron stars act as unipolar inductors, creating very high electric potentials. The acceleration of charged particles
to very high relativistic energies is at the basis of the nonthermal pulsed radiation (and of the hot spots from backflowing particles that we have just seen). The details of the mechanisms, as well as the sites of the acceleration, however, are still uncertain. The particle acceleration may occur near the surface, or in the outer magnetosphere. Goldreich and Julian \cite{Goldreich1969} first  observed that neutron stars are unlikely to be surrounded by vacuum, as the induced electric force has a component parallel to the magnetic field that wins over the gravitational force at the star surface (by many orders of magnitude). 
The charge density in the neutron-star magnetosphere has a maximum value of $\rho_{\textrm{GJ}}\approx-\vec{\Omega}\vdot\vec{B}/(2\pi c)$ (Goldreich--Julian charge density) which screens the electric field parallel to the magnetic field; the particle and the magnetic field corotate with the neutron star possibly up to the so-called `light cylinder', with radius $R_{\mathrm{LC}}=cP_{\textrm{spin}}/(2\pi )$, the imaginary surface where the azimuthal velocity of the (co-rotating) magnetic field reaches the speed of light (Fig.\,\ref{fig:crab_magnetosphere}). Therefore, there are magnetic field lines that close within $R_{\mathrm{LC}}$, and open field lines that cross the light cylinder. The regions of the star where the open field lines start are the \emph{polar gaps}. The induced potential drop along the open field lines is $\Delta V\approx6\times10^{12}B_{12}/P_{\textrm{spin}}^2$\,V, and the lines and particles that flow along them become the pulsar wind. The particle moving along the field lines emit via synchrotron and curvature radiation (the latter is believed to be the main radiation mechanism at GeV energies), and may Compton scatter the soft thermal photons from the neutron star at higher energies. Photons $>$1\,GeV may materialize in the strong field near the surface of the star producing pairs ($\gamma + B \rightarrow e^+ + e^-$) or starting cascades of pairs. The surviving pairs may be responsible for the hot spots and/or contribute to the coherent radio emission process. The existence of bright gamma-ray pulsars, and the fact that their gamma-ray peaks are often not in phase with the radio pulses led the astrophysicists to consider other acceleration sites in the magnetosphere, near the light cylinder where the magnetic field decreases as $r^{-3}$. Here, the magnetic field is much fainter and therefore less high-energy photons are lost in the pair production; these emission models are called \emph{outer-gap} models. Other possible acceleration sites are the \emph{slot gaps} along the last open magnetic field lines (the edges of the polar gaps). Moreover, reconnection in the striped wind outside the light cylinder has been envisioned as an additional feasible mechanism; for an oblique rotator, the asymptotic wind magnetic field
near the rotational equator should consist of stripes of a toroidal field with alternating polarity. In general, these different models may co-exist in the same pulsar or in pulsars at different evolutionary states.

\begin{figure}
\centering
\includegraphics[width=0.7\columnwidth, trim = 0cm 10cm 8cm 0cm, clip]{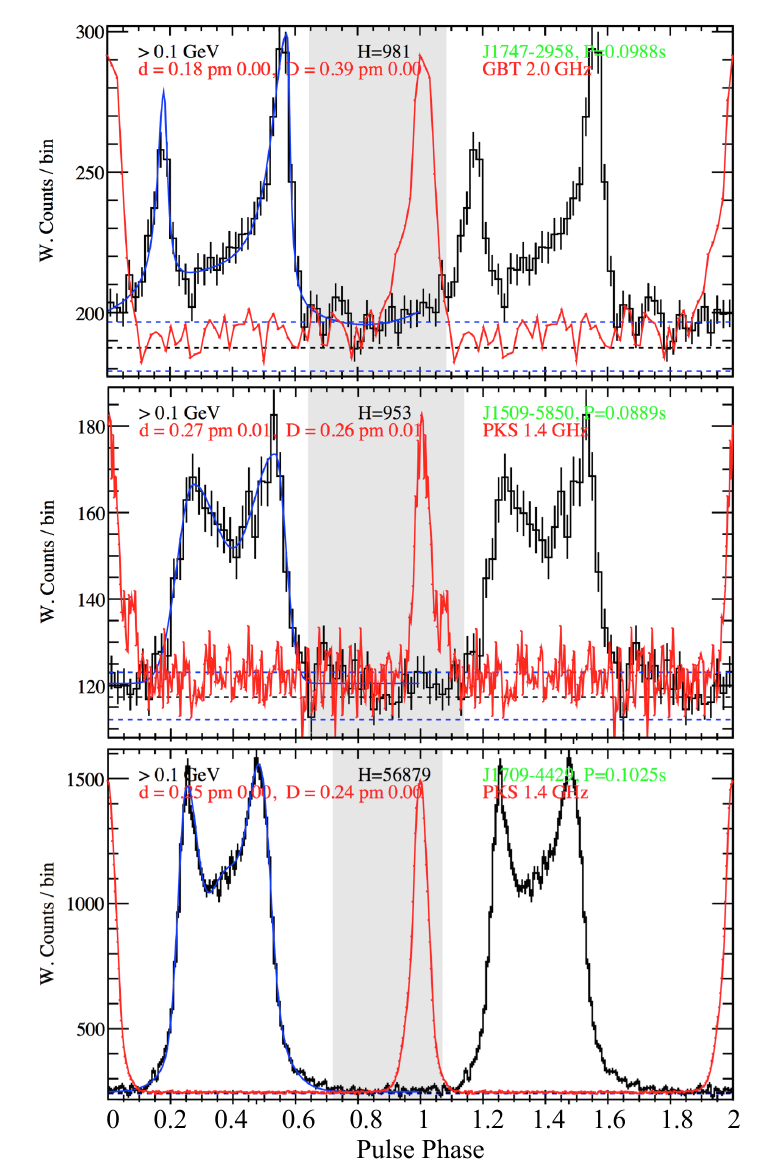} 
\caption{Gamma-ray (black, \emph{Fermi}/LAT data) and radio (red, Parkes or Green Bank telescopes data) pulse profiles for the pulsars J1747--2958 (top), J1509--5850 (middle), J1709--4429 (bottom). From \cite{Abdo2013}.\label{fig:radio_gamma_profiles}}
\end{figure}

The spectrum of magnetospheric emission is generally a power law with a cutoff, $\textrm{d}N/\textrm{d}E=K(E/E_0)^{-\gamma}\exp{-(E/E_\mathrm{c})^\zeta}$ with $\gamma\approx2$; in the case of outer or slot gap models, the cutoff is expected to be exponential ($\zeta=1$) with $E_\mathrm{c}$ in the range between \mbox{1--10\,GeV}, while for the polar gap the cutoff should be super-exponential ($\zeta>1$). The high-quality spectra of many pulsars collected with \emph{Fermi} favour the outer magnetosphere as the main site of production of gamma-ray emission, while it is believed that radio emission is connected to the polar gap and the magnetic pair production. Because the geometry of the polar gap models entails a small beam size, these models have troubles explaining the broad pulsations observed at gamma-rays. Most pulsars show two peaks separated by 0.1--0.5 cycles trailing the radio peak; the larger the gamma-ray peak separation, the smaller the lag (see Fig.\,\ref{fig:radio_gamma_profiles}). 
In the case of the recycled millisecond pulsars, all models have difficulty in explaining their radio emission, since their magnetic fields are so weak that the magnetic-induced electron-positron pair production for most of them should not occur even close to the surface in the polar gaps. A possible solution is a multipolar structure of the magnetic field as such a field would have a surface strength much higher than what is inferred from the spin period and its derivative.  

\section{Magnetars}
\label{sec:magnetar}

Among the isolated neutron stars, \hbindex{magnetars} are certainly the most variable with their unpredictable bursting activities observed in the X-rays and $\gamma$-rays on different time scales, ranging from milliseconds to hundreds of seconds, and long-living enhancements of their X-ray persistent luminosity, commonly referred to as outbursts (see e.g., \cite{Esposito21} for a review and references therein). These transient events trigger the all-sky monitors, such as the Burst Alert Telescope on board the {\it Neil Gehrels Swift Observatory} and the Gamma-ray Burst Monitor on the {\it Fermi} mission, allowing the discovery of new magnetars and the detection of bursting activity from known magnetars. At the time of writing (2023 August), the magnetar family comprises about 30 members\footnote{An updated list is maintained at \url{http://www.physics.mcgill.ca/~pulsar/magnetar/main.html}; \cite{Olausen14}.}, residing in our Galaxy at low latitudes in the Galactic plane, with the exception of two sources located in the Magellanic Clouds, SGR\,0526--66 and CXOU\,J0100--7211. 

Magnetars show pulsations at relatively long periods ($P \sim$ 1 -- 12\,s), which slow down on time scales of a few thousand years ($\dot{P} \sim 10^{-13}-10^{-11}$\,s\,s$^{-1}$). The timing parameters imply characteristic ages of $\tau_{\rm c} \sim 10^3 - 10^5$\,yr and external dipolar magnetic fields of $B_{\rm dip} \sim 10^{14} - 10^{15}$\,G, making them the strongest magnets we know of in the Universe. There is a growing evidence that a stronger field might be present inside the neutron star and in non-dipolar components in the magnetosphere. The persistent X-ray luminosity of $L_{\rm X} \sim 10^{32} - 10^{36}$\,\lum\ is generally higher than their spin-down luminosity $\dot{E}$, pointing to an extra source of power. For this reason, it is believed that magnetar emission is fed by the instabilities and decay of their superstrong magnetic field.  

\subsection{Magnetar history in a nutshell}

The dawn of the neutron star class of magnetars dates back to March 5th, 1979, when an intense burst of hard X-/soft $\gamma$-rays was detected in the direction of the Large Magellanic Cloud, from the source currently known as SGR\,0526--66 \cite{Mazets1979a}. This {\it giant flare} showed features different from those of gamma-ray bursts (GRBs): it was much brighter than a usual GRB, and the emission decayed over a time scale of about 200\,s with a tail modulated at a period of 8.1\,s, suggesting a neutron-star origin for this event. The source was found to be recurrent, with emission of short soft $\gamma$-ray bursts in the following months. Meanwhile, more and more of such bursts were observed and initially classified as GRBs. However, they were coming repeatedly from the same direction in the sky and exhibited a softer spectrum than those of most GRBs, hence the source designation as Soft Gamma-ray Repeaters (SGRs). For instance, the source of GRB 790107 was observed to repeat more than 100 times between 1979 and 1986, and nowadays is recognised as SGR\,1806--20. 

In the same years, an unusual 7\,s X-ray pulsar, 1E\,2259$+$586, was discovered as a bright persistent source ($L_{\rm X} \sim 10^{35}$\,\lum) at the center of the supernova remnant G109.1--1.0 \cite{Fahlman81}. Originally thought to be in a binary system with an elusive low-mass companion star, this object shared similar characteristics with a handful of other sources, such as 1E\,1048.1--5937 and 4U\,0142$+$614: bright X-ray pulsations at few-second periods, an X-ray luminosity exceeding the spin-down energy loss rate, and no apparent companion from which to accrete matter. Owing to these traits, these new sources were awarded the epithet of Anomalous X-ray Pulsar (AXP). 

Following the discovery of SGRs as a peculiar group of bursting isolated neutron stars, a variety of models were put forward. The most successful explanation is provided in terms of a highly magnetized neutron star, hence the name {\it magnetar}, with $B_{\rm dip} \sim 10^{14} - 10^{15}$\,G \cite{Duncan92}. According to the magnetar model, the dominant energy reservoir in these sources is magnetic: giant flares are caused by a large-scale reconnection of the magnetic field and the short fainter bursts are triggered when magnetic stresses build up in a portion of the crust sufficiently to crack it \cite{Thompson95,Thompson96}. Thompson \& Duncan \cite{Thompson96} further pointed out similarities between the AXPs and SGRs in their quiescent state, arguing that AXPs are also magnetars with a decaying field powering the high X-ray luminosity. Alternative scenarios considered isolated neutron stars with $\sim 10^{12}$\,G magnetic fields accreting from a fall-back disk formed after the supernova explosion \cite[see e.g.,][]{Alpar2001}, however these models fail to explain the powerful flaring activity observed in these sources.

The confirmation of the magnetar picture was brought by the measurement of the period derivative of an SGR for the first time. Using {\it Rossi X-Ray Timing Explorer} data, Kouveliotou {\it et al.} \cite{Kouveliotou98} discovered pulsations in the persistent X-ray flux of SGR\,1806$-$20 at a period of 7.47\,s and a spindown rate of $2.6 \times 10^{-3}$\,s\,yr$^{-1}$, implying $B_{\rm dip} = 8 \times 10^{14}$\,G and $\tau_{\rm c} =1.5$\,kyr. Moreover, the observed X-ray luminosity was two orders of magnitude higher than the rotational energy loss rate. Therefore, only the magnetar model could account for these observed properties. A few years later, the detection of SGR-like bursts from two AXPs (1E\,2259+586 and 1E\,1048.1$-$5937) blurred the boundary between SGRs and AXPs, observationally unifying the two groups as predicted by Thompson \& Duncan \cite{Thompson96}. Since then, it is clear that both SGRs and AXPs host magnetars, which can show a continuous spectrum of behaviours, from long periods in quiescence to strong bursting activity and major flares.

\subsection{Persistent emission}
\label{subsec:persistent}

\begin{figure}
\centering
\includegraphics[width=0.9\columnwidth, trim = 3cm 0cm 2cm 0cm, clip=True]{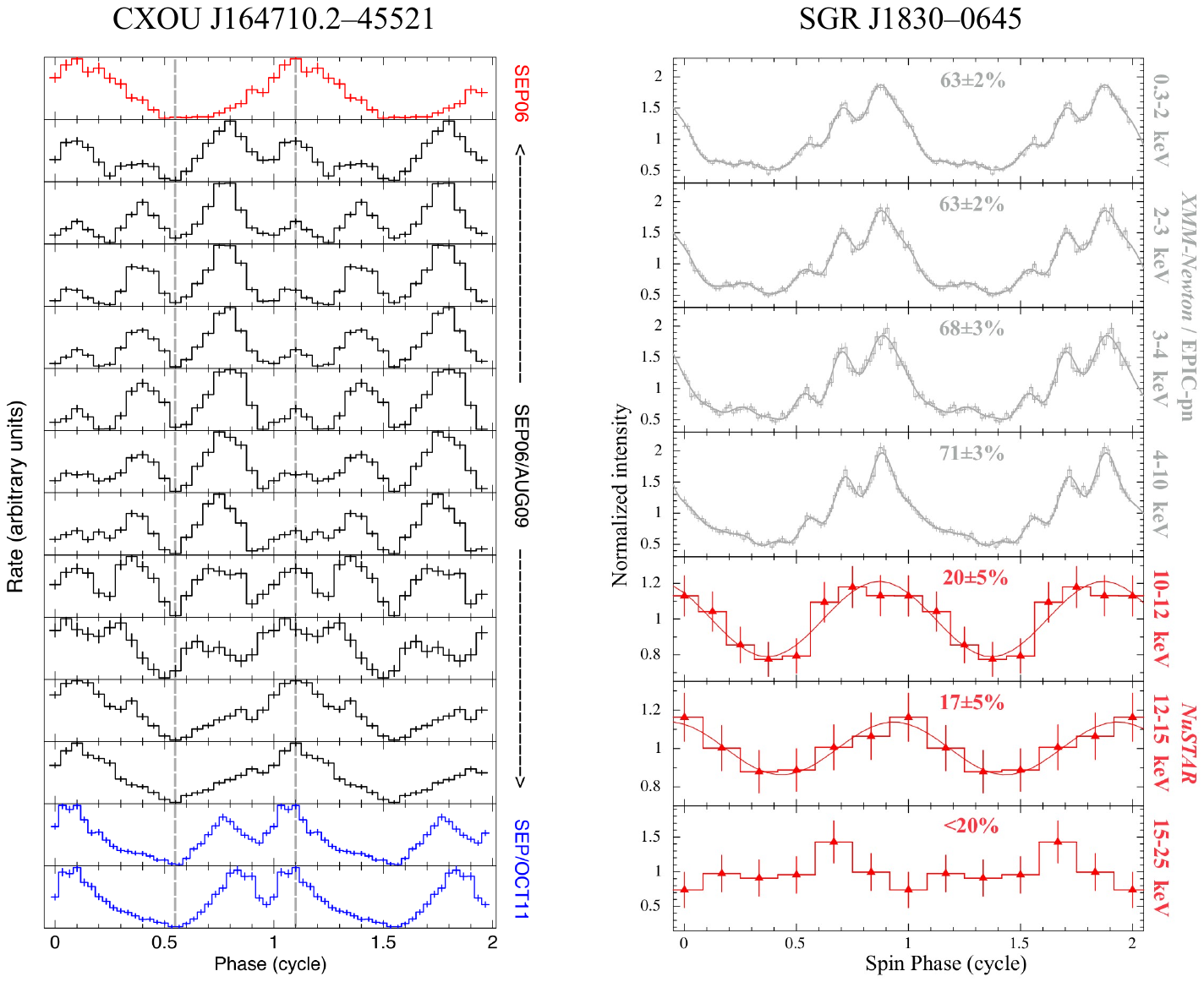} 
\caption{\label{fig:mag_pp} {\it Left:} Multi-epoch soft X-ray pulse profiles of the magnetar CXOU\,J164710.2--45521 from 2006 September in quiescence (red), during the 2006 (black) and 2011 (blue) outbursts. Both the 2006 pre-outburst and 2011 outburst folded light curves have been shifted in phase in order to align their minima with those of the 2006 outburst. From \cite{Rodriguez14}. {\it Right:} Energy-resolved pulse profiles of the magnetar SGR\,J1830--0645 extracted from quasi-simultaneous {\it XMM–Newton}/EPIC-pn (gray) and {\it NuSTAR} (red) observations performed at the outburst peak. The best-fitting models obtained by using seven sinusoidal components (fundamental plus harmonics) for EPIC-pn and a single sinusoidal component (fundamental) for {\it NuSTAR} are indicated with solid lines. The corresponding pulsed fraction values (or the 3$\sigma$ upper limit for the 15--25\,keV range) are reported in each panel. From \cite{Cotizelati21}.} 
\end{figure}

The primary manifestations of this group of highly magnetized neutron stars occur in the X-ray energy range. All confirmed magnetars display pulsations in the soft X-ray band ($<10$\,keV) with pulse profiles modelled with one or multiple sinusoidal functions and pulsed fraction spreading over a wide range. The shape of the pulse profiles are usually energy dependent and time variable: changes are often observed in correspondence with outbursts or strong bursting periods (see Sec.\,\ref{subsec:transient_em}). For instance, the pulse profile of the magnetar CXOU\,J164710.2--45521, hosted within the massive star cluster Westerlund\,I, evolves from a single-peaked structure during the quiescent state to a multi-peaked configuration in outburst (see Fig.\,\ref{fig:mag_pp}, left panel). Discovered on October 10th, 2020, the magnetar SGR\,J1830--0645 displays a pulse profile with a relatively complex structure below 10\,keV, with a wide dip and subpeaks that simplify at higher energies (see Fig.\,\ref{fig:mag_pp}, right panel). 
  
The soft (0.5--10\,keV) X-ray part of the persistent emission spectrum in quiescence is typically well described by a thermal component (a blackbody with temperature $kT \sim 0.3-1$\,keV) and, in some cases, a second component is required, either a hotter blackbody ($kT \sim 1-2$\,keV) or a power law with photon index $\Gamma \sim 2-4$. Since most magnetars are located at low Galactic latitudes, their spectra are heavily absorbed with hydrogen column densities $N_{\rm H}$ ranging between $10^{21}$ and $10^{23}$\,cm$^{-2}$. Note that in the spectral fitting procedure, the $N_{\rm H}$, the power-law photon index, and the blackbody temperature are correlated fit parameters, making it challenging to disentangle and constrain the contribution from the two components. Moreover, the $N_{\rm H}$ values derived from the blackbody+power law models are often higher than those estimated using only thermal components, suggesting that the power-law component diverges at low energy. The magnetar CXOU\,J0100--7211 in the Small Magellanic Cloud has the lowest interstellar absorption ($N_{\rm H} \sim 6 \times 10^{20}$\,cm$^{-2}$) and, hence, offers the best opportunity to examine the X-ray emission at low energies. Its spectrum is well described by a two-blackbody model with temperatures $\sim0.3$\,keV and $\sim0.7$\,keV, while a blackbody+power law fit is rejected \cite{Tiengo08}. It is clear that caution is demanded when drawing physical interpretations from the spectral parameters. The phenomenological decomposition of thermal+power law components is generally interpreted in a scenario according to which the thermal emission that arises from the hot neutron star surface could be distorted by a magnetized atmosphere and magnetospheric effects, such as resonant cyclotron scattering (for a detailed discussion about these models see \cite{Turolla15}). Soft seed photons emitted at the surface are boosted to higher energies due to repeated scatterings with charged particles flowing in a twisted magnetosphere leading to the formation of a power-law tail.  

\begin{figure}
\centering
\includegraphics[width=0.8\columnwidth]{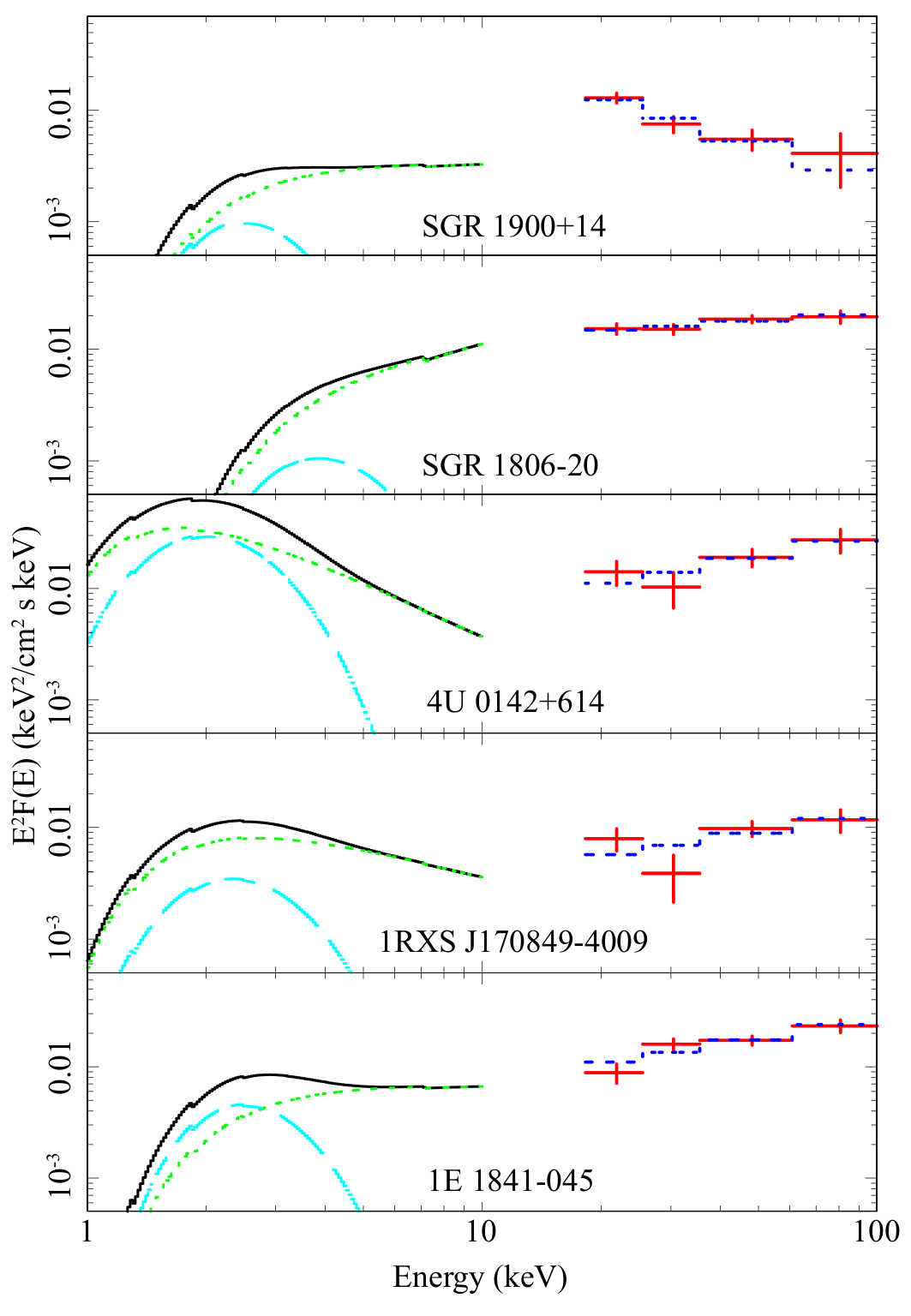} 
\caption{\label{fig:mag_spec_hardX} Broad band X-ray spectra ($1-100$\,keV) for the five magnetars with detected presistent emission at hard X-rays. The red data points above 20\,keV are the {\it INTEGRAL} spectra fitted with a power law (dotted blue line). The solid black line represents the best-fitting model for the soft X-ray ($<10$\,keV) spectrum, which includes a blackbody (dashed light blue line) and a power law (dotted green line). From \cite{Gotz06}.}
\end{figure}

Several magnetars have been detected in the hard X-rays up to $\sim150-200$\,keV. First observed with {\it INTEGRAL} from persistently bright sources and later confirmed by {\it Suzaku} and {\it NuSTAR}, the hard ($>10$\,keV) X-ray component is adequately modelled by a power-law-like spectrum flatter than that observed in the soft X-rays, with photon index $\Gamma \sim0.5-2$. Fig.\,\ref{fig:mag_spec_hardX} shows the broad band spectra for five magnetars with detected hard X-ray persistent emission. 1E\,1841--045 in the supernova remnant Kes\,73 was the first to be detected in the hard X-ray energy range and a reanalysis of archival {\it Rossi X-Ray Timing Explorer} data unveiled pulsations up to $\sim150$\,keV, confirming that the hard X-ray emission orignates from the magnetar and not from the host supernova remnant \cite{Kuiper06}. The hard X-ray component of SGR\,1900+14 was pinpointed while the source was in quiescence and exhibits a steeper spectrum than those of the other magnetars, with photon index $\Gamma \sim3$ \cite{Gotz06}. The hard X-ray component is modulated at the neutron star spin period and its luminosity is comparable to or higher than that measured in the soft X-rays. The hard X-ray pulsed emission usually shows a harder spectrum than that of the averaged hard X-ray emission and may display phase dependent variations, such as peak shifts, and/or changes in the pulse profile shape with energy. As an example, the hard X-ray pulse profile of SGR\,J1830--0645 shown in Fig.\,\ref{fig:mag_pp}, right panel, exhibits a simpler morphology than the profile below 10\,keV. However, the trend of the pulsed fraction is different from what is usually observed in other magnetars: it markedly drops above 10\,keV, where the hard power-law tail dominates the source emission, pointing out that the hard X-ray component is pulsed to a smaller extent than the soft thermal component.

The upper limits derived in the MeV region imply that the hard X-ray spectra have to break/bend between $\sim200$\,keV and 750\,keV. Moreover, stringent upper limits between $10^{-12}-10^{-11}$\,\flux\ were derived in the $0.1-10$\,GeV interval employing 6 years of {\it Fermi} Large Area Telescope observations, and searches for emission above 100\,GeV revealed to be unsuccessful. The mechanism at the origin of the hard X-ray tails is still under debate. They might be due to either bremsstrahlung photons emitted by the neutron star surface layers heated by returning currents, or synchrotron emission from pairs produced at a height of about 100 km above the neutron star \cite{Thompson05}. However, the prime candidate is resonant Compton up-scattering of soft thermal photons by a population of highly relativistic electrons threaded in the magnetosphere \cite[see e.g.,][]{Wadiasingh18}.

Investigating magnetar emission at the optical and infrared bands is challenging, due to their intrinsic faintness at these wavelengths, and to their locations in strongly absorbed regions of the Galactic plane. Nonetheless, counterparts have been found for about two third of them. The associations are unambiguous for three sources since optical pulsations at the X-ray period have been measured \cite[see e.g.,][]{Dhillon11}. All three objects (4U\,0142+61, 1E\,1048.1--5937, and  SGR\,0501+4516) present optical light curves with broadly similar pulsed fractions and pulse profiles that are approximately in phase with the X-rays ones. For 4U\,0142+61 and 1E\,1048.1--5937, the optical and X-ray pulse profiles show similar morphology. For SGR\,0501+4516, while the folded optical and 8--12\,keV X-ray light curves share a double-peaked shape, the 0.5--10\,keV X-ray one exhibits a single-peaked profile. The optical/infrared emission can be explained within the magnetar scenario by non-thermal emission from the inner magnetosphere, where the emission at these wavelengths is produced by curvature radiation emitted by relativistic electrons moving along the closed field lines \cite{Zane11}. However, a detailed model is still lacking. 

\subsection{Transient emission}
\label{subsec:transient_em}

Transient activity is the birthmark of magnetars. Their radiative variability includes short, explosive events from milliseconds to hundreds of seconds (i.e., bursts and giant flares) and longer-lived outbursts (weeks to months), i.e., increase of the persistent X-ray luminosity up to three orders of magnitude with respect to the quiescent level. In the following, we discuss these three phenomena in detail.\\

\begin{figure}
\centering
\includegraphics[width=0.9\columnwidth, trim = 0cm 8cm 0cm 0cm, clip]{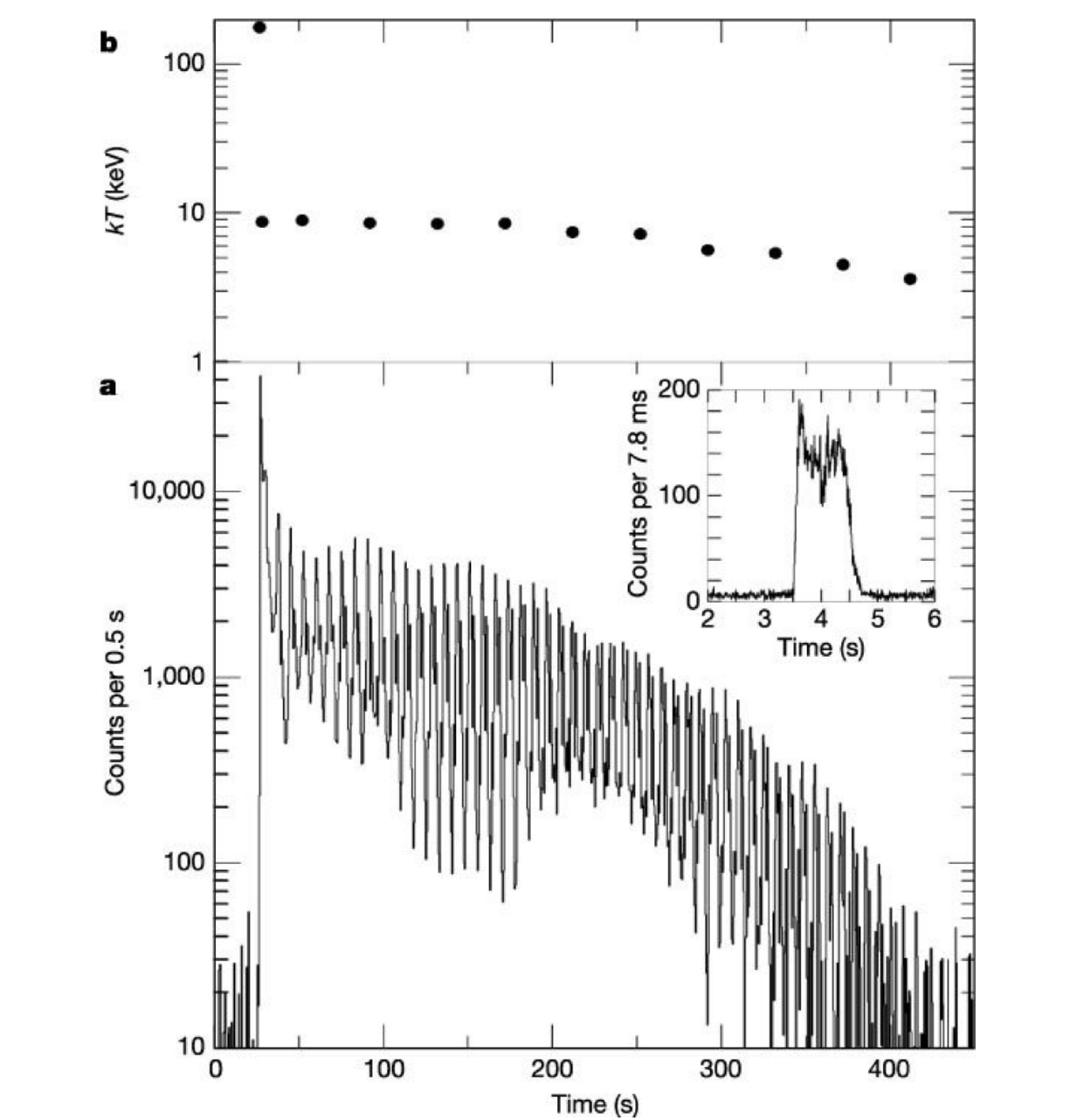} 
\caption{\label{fig:mag_gf} The 27 December 2004 giant flare of SGR\,1806--20. {\it Panel a}: 20--100\,keV light curve with 0.5\,s time resolution, extracted from {\it RHESSI} data and showing the initial spike (which saturated the detector) at $\sim 26$\,s. The inset shows the light curve of the precursor with 8\,ms time resolution. {\it Panel b}: Temporal evolution of the blackbody temperature. From \cite{Hurley05}.}
\end{figure}

{\it Giant flares.} 
So far, only three giant flares from three different sources have been recorded: in 1979 from SGR\,0526--66 \cite{Mazets1979a}, in 1998 from SGR\,1900+14 \cite{Hurley99}, and in 2004 from SGR\,1806--20 \cite{Hurley05}. The detection of only three of such events in about 40 years implies that these events are rare. The properties of the three giant flares are very similar: they began with a brief ($\sim 0.1-0.2$\,s) spike of $\gamma$-rays, with emission detected up to a few MeVs that reached a peak luminosity $\geq 10^{44}-10^{45}$\,\lum\ for SGR\,0526--66 and SGR\,1900+14, and $\geq 10^{47}$\,\lum\ for SGR\,1806--20. The initial flashes were followed by hard X-ray tails strongly modulated at the spin period of the neutron stars and observed to decay in a few minutes. Interestingly, the energy released in the pulsating tails was comparable for all the three events ($\sim 10^{44}$\,erg). According to the magnetar model, all the energy is released during the initial spike, when a hot fireball is launched. Part of this energy is trapped in regions of closed magnetic field lines in the magnetosphere and is converted into a photon-pair plasma, which cools down producing the radiation observed in the oscillating tail \cite{Thompson95}. The fraction of the trapped energy depends on the strength of the magnetic field. Therefore, the fact that this quantity is similar in the three giant flares is consistent with the three magnetars having magnetic fields of the same order of magnitude. Fig.\,\ref{fig:mag_gf} displays the temporal evolution of the intensity and the temperature of the giant flare of SGR\,1806--20. The initial spike showed a spectrum consistent with a blackbody with an average temperature $kT \sim 200$\,keV, while the time-resolved spectra of the modulated hard X-ray tail were well fit by both a thermal bremsstrahlung and blackbody model. However, a softening of the emission is clear with the temperature decreasing to $\sim 10$\,keV immediately after the initial spike. Two distinctive traits made this event slightly different from the two previously observed: a $\sim$1-s-long precursor was detected about 140\,s before the flare with a roughly flat-topped profile (see Fig.\,\ref{fig:mag_gf} inset), whose spectrum can be approximated by a thermal bremsstrahlung with an energy of $\sim 5 \times 10^{42}$\,erg. The peculiar light curve and large energy output distinguished the precursor from a usual magnetar-like short burst. Moreover, this event was characterised by a higher spike-to-tail energy ratio ($\geq100$) than the two other giant flares ($\sim1$).

A radio counterpart, in the form of a transient extended radio emission, was detected following the giant flares of SGR\,1900+14 and SGR\,1806--20 \cite{Frail99, Gaensler05}. For the latter, the radio afterglow reached a luminosity 500 times larger than that seen in SGR\,1900+14 radio nebula and was found to be linearly polarized, implying that the emission mechanism is synchrotron radiation. At least $\sim 10^{43}$\,erg of energy must have been emitted to create such radio nebula during the giant flare, in the form of relativistic particles and magnetic fields. Another remarkable phenomenon detected during the giant flares is the occurrence of quasi-periodic oscillations in the pulsating tails (see e.g., \cite{Israel05}). Only detected in specific rotational phase intervals, these modulations are believed to be due to seismic vibrations of the neutron star induced by the giant flare.

At extragalactic distances, only the brief initial hard spike of a giant flare would be detectable and would resemble a short gamma-ray burst. Two possible candidates for extragalactic magnetar giant flares, GRB\,070201 in M31 and GRB\,051103 in M81, had been proposed due to their energy output, localisation, short duration, and hard spectrum. 
Recently, GRB\,200415A was identified as the third and most promising extragalactic giant flare \cite{Fermi21}. On April 15th, 2020, the {\it Fermi} Gamma-ray Burst Monitor triggered on a short gamma-ray burst, coming from a direction compatible with the nearby Sculptor galaxy at $\sim 3.5$ megaparsecs. The {\it Fermi} Large Area Telescope detected GeV emission from $\sim 20$\,s until $\sim 5$\,min after the main event. The long delay of $\sim 20$\,s between the trigger and the detection of the hard $\gamma$-ray afterglow sets this event apart among the short gamma-ray burst population collected by {\it Fermi}. Although this property does not rule out a cosmological short gamma-ray burst nature for this event, its association with the Sculptor galaxy, the very flat spectrum below 1\,MeV, and the discovery of a quasi-periodic oscillation in the main peak of the light curve \cite{CastroTirado21} strongly point toward a magnetar giant flare interpretation. This recent finding prompted a search for new candidate extragalactic giant flares using the largest available sample of gamma-ray bursts and new galaxy catalogs. Four additional candidates were singled out by virtue of their rise time, isotropic energy and prompt emission, which are inconsistent with the properties of cosmological gamma-ray bursts \cite{Burns21}. However, the peculiarity of the known magnetar giant flares is the long periodic tails, which are modulated at the magnetar spin period and show quasi-periodic oscillations. Unfortunately, these signatures are not unambiguously identifiable at extragalactic distances with
existing instruments.\\

{\it Short Bursts.}
Short bursts are the benchmark of the magnetar transient activity, representing the main way to discover new magnetars, and they often mark  the onset of a new outburst. They can occur sporadically or clustered in time, and it is not possible to predict their appearance and which source is about to burst. There are magnetars that appeared dormant for decades and suddenly entered an active phase with the emission of hundreds of bursts in a few days. Burst durations span over two orders of magnitude, ranging from a few milliseconds to a few seconds, and the peak luminosities is found in the range $\sim 10^{36}-10^{43}$\,\lum. Their light curves are single- or multi-peaked, with a rise time typically faster than the decay time. The brightest bursts, also called intermediate bursts, sometimes show a tail lasting up to hours that might contain more energy than the initial spike. Burst-to-tail energy ratios can vary by one order of magnitude in different magnetars and even in different bursts of the same source. Magnetar bursts typically have no emission above $\sim 200$\,keV. Broad band spectra are described by different models, including a simple blackbody, double blackbodies, optically thin thermal bremsstrahlung models, or Comptonized models (i.e., a power law with an exponential cutoff at higher energies). More than one model usually gives a satisfactory description of the burst spectrum, preferring the single- or double-blackbody model. The blackbody temperature $kT$ is generally between $\sim 2$ and 12\,keV, and when a second thermal component is required, a higher temperature blackbody ($kT \sim 12-13$\,keV) with a smaller radius shows up besides the lower temperature one.  

Several statistical studies of the properties of short bursts as a population have been performed. For instance, a catalog covering the first five years of {\it Fermi} Gamma-ray Burst Monitor of magnetar burst observations was compiled \cite{Collazzi15}. This comprehensive overview of about 450 bursts highlighted similarities in the spectral and temporal parameters of the flaring events regardless the source. The distribution of the $T_{\rm 90}$, defined as the time during which the cumulative counts rise from 5\% to 95\%, tends to center around $\sim 100$\,ms for the events from known magnetars. Regarding the spectral properties, the Comptonized model parameters are similar among the different bursts with a peak energy centered at $\sim 40$\,keV, while a two-blackbody model yields temperatures centered around $\sim4.5$ and $\sim15$\,keV throughout the sample. Furthermore, for a given source and instrument, the number $N$ versus fluence $S$ distribution of bursts follows a power-law function ($N$($>S$) $\propto S^{-\alpha}$) with indexes in the range $\sim 0.6-0.9$.

\begin{figure}
\centering
\includegraphics[width=0.95\columnwidth, trim = 0cm 0cm 8cm 0cm, clip]{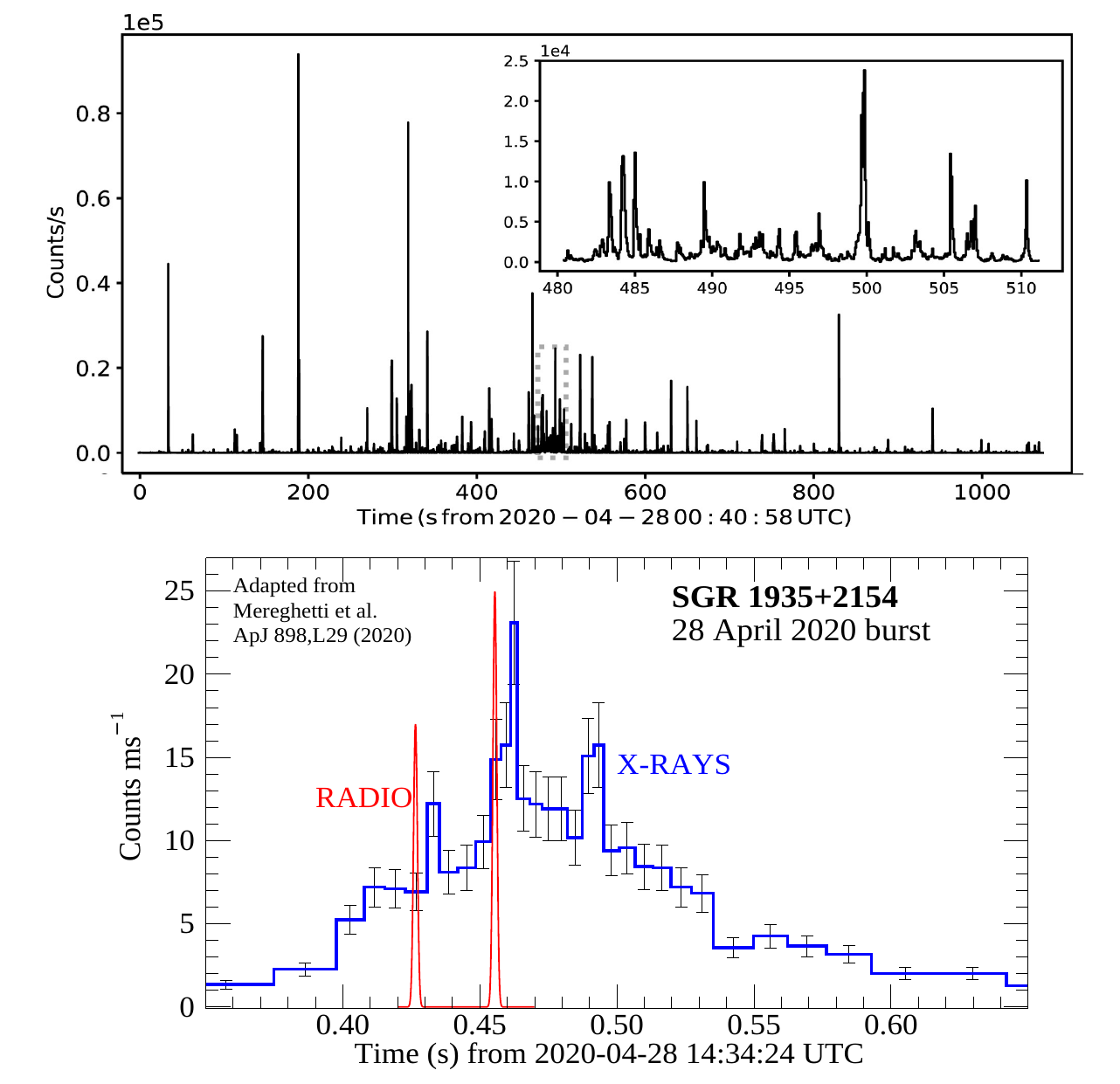} 
\caption{\label{fig:mag_burst} {\it Top panel}: {\it NICER} light curve of the burst forest emitted by SGR\,1935+2154 in 2020 April. The light curve was extracted from the first $\sim 1200$\,seconds of the observation acquired on 2020 April 28, starting at 00:40:58 UTC, in the $1 - 10$\,keV energy range. More than $\sim 220$ bursts were detected. The inset is a zoom-in at the area delimited with a dotted gray box, representing the most intense bursting period. From \cite{Younes20}. {\it Bottom panel}: 
Light curves of the simultaneous X-ray (blue) and radio (red) burst, as observed with the {\it INTEGRAL} IBIS/ISGRI instrument and CHIME respectively, emitted by SGR\,1935+2154 on 2020 April 28. Adapted from \cite{Mereghetti20}.}
\end{figure}

SGR\,1935+2154 is one of the most burst prolific magnetars. Since its discovery in 2014, it emitted about 300 bursts, as solitary events or associated with outbursts, and a burst forest (i.e., emission of a plethora of short bursts clustered in time) on April 27th-28th, 2020. The average burst energy increased during the four episodes registered between 2014 and 2016, suggesting that the next bursting activity would have likely been more intense. Contrary to this expectation, the burst energies of the 2019 and 2020 episodes (excluding the burst forest for the latter) indicate a flattening of the average burst energy curve. Moreover, the bursts detected during the latest two activations were slightly longer and softer than those observed at earlier epochs \cite[see e.g.,][]{Lin20b}. On April 27th, 2020, SGR\,1935+2154 entered a new exceptional active phase, whose onset was marked by a storm of highly energetic bursts detected with the {\it Swift} Burst Alert Telescope and the {\it Fermi} Gamma-ray Burst Monitor. {\it NICER} observed the source $\sim 6$ hours after the initial trigger and caught the tail of the burst forest (see Fig.\,\ref{fig:mag_burst}, top panel; \cite{Younes20}). During the first $\sim 20$ minutes of the pointing, more than 220 bursts were identified translating into a burst rate of $>0.2$\,burst\,s$^{-1}$, to be compared with the rate of 0.008\,burst\,s$^{-1}$ derived three hours later. The majority of the bursts exhibited multi-peaked profiles with shorter rise than fall times, in agreement with the bulk of short magnetar bursts, and the $1 - 10$\,keV spectra are well modelled by an absorbed blackbody with an average temperature of 1.7\,keV. However, what made this new active phase really unique is the detection of a bright, millisecond-duration radio burst, with properties reminiscent those of fast radio bursts. This was the first time to observe such an event from a know magnetar. A day after the initial trigger, the Canadian Hydrogen Intensity Mapping Experiment (CHIME) and the Survey for Transient Astronomical Radio Emission 2 (STARE2) independently observed the radio burst \cite{Chime20, Bochenek20}. The energy released $E_{\rm radio} \sim 10^{34}-10^{35}$\,erg is about three orders of magnitude greater than that of any radio pulse from the Crab pulsar, previously the source of the brightest Galactic radio bursts, and 30 times less energetic than the the weakest extragalactic fast radio burst observed so far. Remarkably, the radio burst was temporally coincident with a hard X-ray burst \cite[see e.g.,][]{Mereghetti20}, showing for the first time that magnetar bursts can have a bright radio counterpart. The bottom panel of Fig.\,\ref{fig:mag_burst} shows the light curves of the hard X-ray burst as observed by {\it INTEGRAL} and of the double-peaked radio burst detected by CHIME. The X-ray light curve consists of a broad pulse with three narrow peaks separated by $\sim 30$\,ms. The X-ray peaks lag the radio ones by $\sim 6.5$\,ms; this short delay implies that both components arise from a small region of the neutron star magnetosphere. The X-ray burst spectrum was harder than any other extracted from bursts emitted by the same source, while the energy output $E_{\rm X} \sim 10^{39}$\,erg is in the range of energies expected from magnetar bursts, resulting in $E_{\rm radio} / E_{\rm X} \sim 10^{-5}$. This discovery motivated the search for bright radio bursts in archival radio and X-ray simultaneous data of magnetar outbursts. Another detection was reported during the 2009 outburst of the magnetar 1E\,1547.0--5408 \cite{Israel21}. Two radio bursts were observed, and one of them was anticipated by a short X-ray burst by $\sim1$\,s, implying $E_{\rm radio} / E_{\rm X} \sim 10^{-9}$. These two detections strengthened the hypothesis that at least a sub-group of fast radio bursts can be powered by magnetars at cosmological distances. Furthermore, the wide range of the radio-to-X-ray energy ratios suggests that magnetar radio bursts can resemble both the powerful fast radio burst and the typical singles pulses from ordinary radio pulsars, bridging the two populations. \\

\begin{figure}
\centering
\includegraphics[width=0.98\columnwidth]{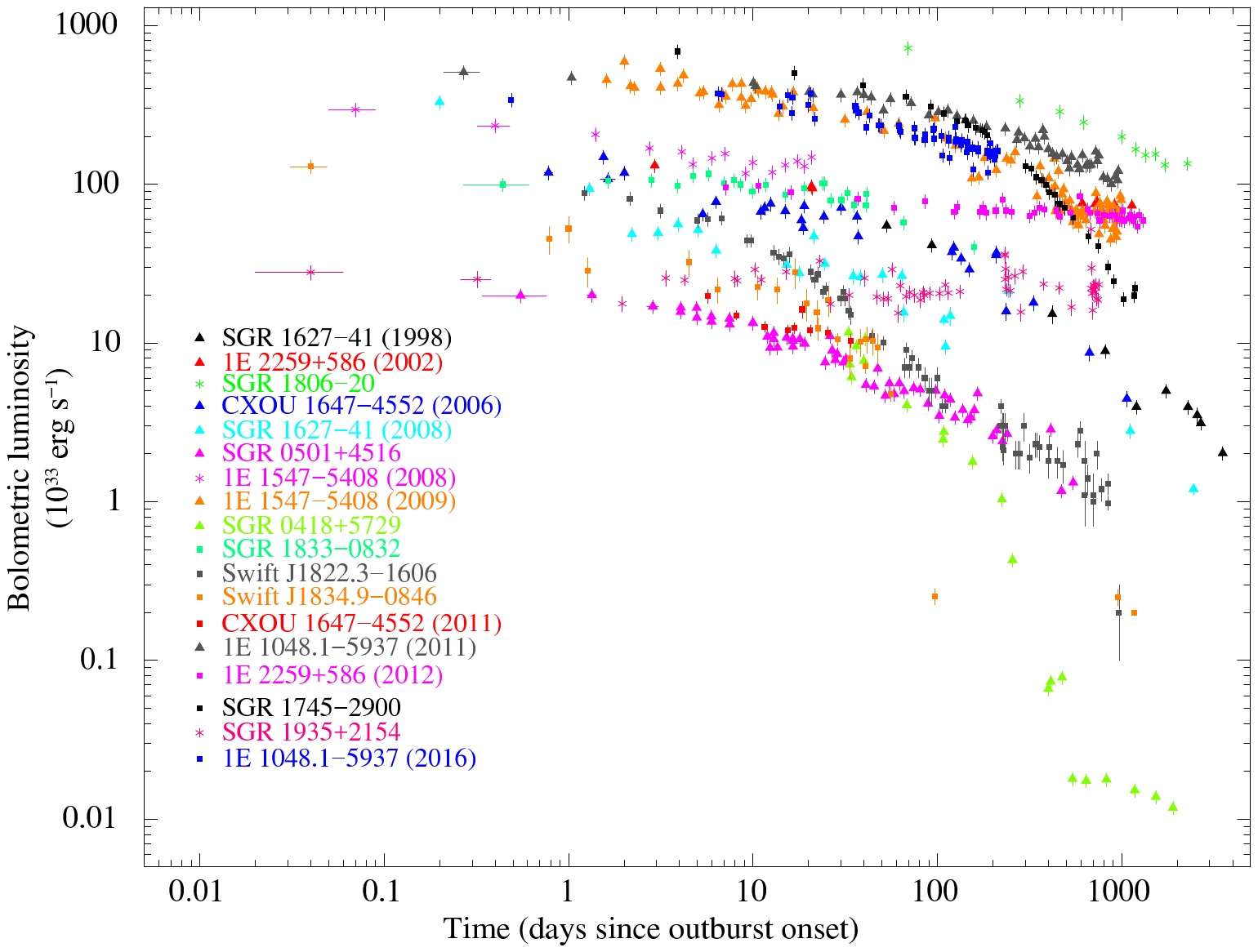} 
\caption{\label{fig:mag_mooc} Temporal evolution of the bolometric ($0.01-100$\,keV) luminosities for the major outbursts occurred up to the end of 2016 and with intende and long coverage. From \cite{Cotizelati18}.}
\end{figure}

{\it Outbursts.}
Short bursts are often accompanied by large enhancements of the persistent X-ray flux that can increase up to three orders of magnitude higher than the pre-outburst level. Then, the flux usually relaxes back to the quiescent level on time scales ranging from weeks to months/years.
The decay pattern varies from outburst to outburst, but in most cases is characterised by a very rapid initial decay within minutes to hours, followed by a slower fading modelled by a power-law or exponential function. Sometimes, the flux decrease might be interrupted by a period of stability or can drop suddenly. Fig.\,\ref{fig:mag_mooc} shows the long-term light curves of all the outbursts detected up to the end of 2016 and monitored intensively in X-rays.

During an outburst, the X-ray spectrum undergoes an overall initial hardening and then slowly softens on the time scale of the flux relaxation. For instance, for a soft X-ray spectrum modelled with an absorbed blackbody and a power law, the hardening may correspond to an increase in the blackbody temperature and a decrease in the photon index. While, if only one thermal component is required for the quiescent spectrum, a second hotter blackbody may appear during an outburst. The emission area, temperature and luminosity associated with this new component typically decline as the outburst progresses until becoming undetectable again. In some cases, a hard non-thermal tail is detected at the outburst peak and observed to fade on faster time scales than the softening of the $0.3 - 10$\,keV X-ray spectrum. 

It is believed that outbursts are caused by heat deposition in a restricted area of the magnetar surface, however the responsible heating mechanism is still poorly understood. The energy is injected into the crust and then conducted up to the surface because of magnetic stresses, resulting in displacements/fractures of the crust itself (see e.g., \cite{Gourgouliatos21}). The energy should be injected into the outer crust, otherwise most of it would be radiated in the form of neutrinos. Moreover, a minimum value of energy is required to yield an observable effect. For energy $< 10^{40}$\,erg, the luminosity increase would not be detectable due to the sensitivity of current satellites. On the other hand, for energy $>10^{43}$\,erg, a saturation effect occurs, meaning that a larger amount of energy does not change the observable result. The surface photon luminosity reaches a limiting value of $\sim 10^{36}$\,\lum\ since the crust becomes so hot that nearly all the energy is released via neutrinos before it reaches the surface \cite{Pons12}. Moreover, the crustal displacements induce a strong twist on the magnetic filed lines outside in the magnetosphere \cite{Beloborodov09}. Additional heating of the surface layers is then produced by the currents flowing in the bundle as they hit the star. The twist must decay in order to supply its own currents; the gradual untwisting induces a reduction of the area impacted by the magnetospheric charges and the luminosity decreases. Both mechanisms are most likely at work during an outburst. In these terms, it is worthy to mention the case of the magnetar 1E\,1547--5408 \cite{Cotizelati20}. About one year after the latest outburst in 2009, the source settled in a relatively high-flux state, which was observed to be steady over the last 10 years. During these 10 years, the soft X-ray flux attained a value $\sim 30$ times higher than its historical minimum measured in 2006. Moreover, hard X-ray observations carried out in 2016 and 2019 revealed a faint emission up to $\sim 70$\,keV, described by a flat power law component with a $10-70$\,keV flux $\sim 20$ times smaller than that at the peak of the 2009 outburst. These properties are at variance with the typical overall softening observed in an ordinary magnetar outburst and can be naturally accounted for by invoking the untwisting bundle scenario as the only mechanism driving the outburst. The source has reached a new persistent magnetospheric state, different from that observed during the pre-outburst epoch. A magnetospheric reconnection can lead to a local reorganization of the magnetic field, leaving a new pattern of twisted lines associated to current bundles that could dissipate over decades. The flowing currents are responsible for both the non-thermal X-ray emission, via resonant Compton scattering of photons from the surface by the charged particles, and the thermal X-ray emission, via the current dissipation in localised regions close to the stellar surface. The 2009 outburst and the new persistent state of 1E\,1547--5408 provide compelling evidence that a small sample of magnetar outbursts can be entirely powered by the dissipation of magnetospheric currents. Moreover, the source underwent a new outburst in 2022 April, its first since its major 2009 outburst, and the flux decayed back to the same pre-outburst new persistent level.

Changes in the timing properties have been reported several times along with outburst episodes. In some cases, the pulse profiles become more complex in shape with a multi-peaked configuration and, accordingly, the pulsed fraction varies (see Fig.\,\ref{fig:mag_pp}, left panel). Glitches, i.e. sudden increases in the rotational frequency of the star, are very common during outbursts with amplitude in the range $\delta \nu / \nu \sim 10^{-9}-10^{-5}$. Note that many glitches are also observed in the absence of measurable change in the X-ray flux. A statistical study of these timing anomalies in magnetars and rotation-powered was performed \cite{Fuentes17}. The glitch activity, defined as the time-averaged change of the rotation frequency due to glitches, of magnetars with the smallest characteristic ages $\tau_{\rm c}$ is lower than that of pulsars with similar $\tau_{\rm c}$. However, their activity is larger than that of pulsars of equal spin-down power. The only parameter for which the glitch activity of
magnetars appears to follow the same relation as for rotation-powered pulsars is the spin-down rate. An interesting aspect of magnetar glitches is the over-recovery following the timing event, often resulting in a net spin-down or anti-glitches. The most famous anti-glitch candidate was reported for the magnetar 1E\,2259+586, whose spin frequency changed by a factor of $\sim -5 \times 10^{-8}$\,Hz in less than four days \cite{Archibald13}. This event was accompanied by multiple X-ray radiative episodes, such as the emission of a short hard X-ray burst and an increase of the persistent flux by a factor of 2. The extreme and rapid variability in the spin-down torque appears to be another common feature following magnetar outbursts. Several magnetars have been intensively monitored after a flaring episode and, when the torque variations are observed, they can dominate the long-term evolution of these sources. 

Radio transient pulsed emission has been detected in a handful of magnetars, associated with an X-ray outburst. XTE\,J1810--197, which briefly became the brightest radio pulsar at the time of this discovery, was the first radio-loud magnetar to be discovered \cite{Camilo06}. 
The radio emission was observed one year after the onset of the X-ray outburst in 2003. The X-ray flux reached the pre-outburst level in early 2007, although the source remained radio loud until late 2008. Radio pulsations were then caught from 1E\,1547--5408 \cite{Camilo07}, PSR\,1622--4950, the only magnetar discovered in the radio band so far \cite{Levin10}, and SGR\,J1745--2900, found at an angular separation of 0.1\,pc from Sgr A$^*$ \cite{Eatough13}. The most recent additions to this small group are Swift\,J1818.0--1607 (see e.g., \cite{Lower20}), discovered in 2020 March, and SGR\,1935+2154, whose radio pulsed emission was detected after the emission of a few clustered short radio bursts \cite{Zhu23}. The spectrum of the pulsating radio emission of magnetars is flatter than that of rotation-powered pulsars: $S\propto\nu^{-0.5}$ for the former and $S\propto\nu^{-1.8}$ for the latter, where $S$ is the flux density and $\nu$ is the frequency. The only exception is Swift\,J1818.0--1607, which shows a steeper spectrum ($S\propto\nu^{-2.3}$) compared with the other radio-loud magnetars. Moreover, the radio emission displays large pulse-to-pulse variability with pulse shapes that can vary considerably on timescales of minutes. The single pulses are usually comprised of narrow, spiky sub-pulses with a high degree of linear polarization.

A systematic study of all outbursts detected in the X-rays up to the end of 2016 was performed by Coti Zelati {\it et al.} \cite{Cotizelati18} who reanalysed about 1100 X-ray observations in a consistent way, focusing on the temporal evolution of the soft X-ray spectral parameters\footnote{This study is complemented by the Magnetar Outburst Online Catalog, \url{http://magnetars.ice.csic.es/}.}. This work highlighted some common trends in all the outbursts by investigating possible (anti-)correlations between different parameters, e.g., peak and quiescent luminosity, decay time scale, energy released during the outburst, $\tau_{\rm c}$, and $B_{\rm dip}$. For instance, they found that a larger luminosity at the outburst onset corresponds to a larger amount of energy released during the outburst, and the outbursts with a longer overall duration are also the more energetic ones. An anti-correlation between the quiescent luminosity and the outburst luminosity increase was discovered, suggesting a limiting luminosity of $\sim 10^{36}$\,\lum\ for the outbursts regardless of the quiescent luminosity level (see \cite{Pons12} and above). The energy released in an outburst linearly depends on $B_{\rm dip}$, and a sort of limiting energy as a function of age is observed. In other words, the young magnetars undergo more energetic outbursts than older ones. These characteristics are naturally explained in terms of field decay.

\subsection{Low-magnetic field magnetars}
\label{subsec:lowb_mag}

The picture according to which magnetar activity is driven by a super-strong magnetic field has been challenged by the discovery of three magnetars with a magnetic field within the range of those of ordinary radio pulsars, $B_{\rm dip} \sim (0.6-4) \times 10^{13}$\,G (see \cite{Turolla13} for a review and reference therein). We should bear in mind that we are referring to the strength of the large-scale, dipolar component of the magnetic field estimated from the spin parameters far away from the stellar surface (see Eq.\,\ref{eq:Bdip}) and that magnetar activity (i.e. bursts and outbursts) is driven by ultra-strong magnetic fields, `hidden' in the star interior and/or at the surface.  

On June 5th, 2009, the detection of a couple of short hard X-ray bursts heralded the existence of a new magnetar, SGR\,0418+5729. Follow-up observations in the soft X-ray band disclosed a bright persistent counterpart with an observed flux of a few $10^{-11}$\,\flux\ and a spin period of $\sim 9.1$\,s. The first 5 months of monitoring yielded only an upper limit on the spin-down rate, $|\dot{P}| < 1.1 \times 10^{-13}$\,s\,s$^{-1}$ (at 3$\sigma$ confidence level), that corresponds to an upper limit on $B_{\rm dip}$ of $3 \times 10^{13}$\,G. This value was the lowest spin-inferred magnetic field among the magnetar population at that time. An unambiguous measure of $\dot{P}$ required more than 3 years of continuous monitoring. A coherent timing analysis with a baseline of $\sim 1200$ days gave $\dot{P}=4(1) \times 10^{-15}$\,s\,s$^{-1}$, implying $B_{\rm dip} \sim 6 \times 10^{12}$\,G and $\tau_{\rm c} \sim 36$\,Myr, giving confirmation that SGR\,0418+5729 was a low-$B$ magnetar. Swift\,J1822.3--1606 was discovered through the detection of a series of magnetar-like bursts on  July 14th, 2011. The fast slew of the {\it Swift} X-ray Telescope promptly detected a bright source at an observed flux level of $\sim 2 \times 10^{-10}$\,\flux\ and pulsating at $\sim 8.4$\,s. {\it ROSAT} serendipitously observed the region of the sky covering the magnetar position in 1993 September. A reanalysis of these observations unveiled a source at a location consistent with that of Swift\,J1822.3--1606 at a flux level of $\sim 4 \times 10^{-14}$\,\flux, most likely being the magnetar in its quiescent state. A campaign lasting 1.3 years allowed for the determination of the period derivative, equal to $\dot{P}=1.34(1) \times 10^{-13}$\,s\,s$^{-1}$ \cite{Rodriguez16}. Therefore, from the timing parameters, $B_{\rm dip} \sim 3 \times 10^{13}$\,G and $\tau_{\rm c} \sim 1$\,Myr. Swift\,J1822.3--1606 is thus the magnetar with the second lowest magnetic field. The last member of this small group is 3XMM\,J185246.6+003317 (3XMM\,J1852; \cite[see e.g.,][]{Zhou14}). It was serendipitously discovered while undergoing an outburst in 2008 during an {\it XMM-Newton} campaign of the supernova remnant Kes\,79. 
The magnetar rotates at a period $P \sim 11.6$\,s. No spin-down was detected during the seven months of the outburst decay. The $3\sigma$ upper limit for the period derivative $|\dot{P}| < 1.4 \times 10^{-13}$\,s\,s$^{-1}$ translates to $\tau_{\rm c} > 1.3$\,Myr and $B_{\rm dip} < 4 \times 10^{13}$\,G, defining 3XMM\,J1852 as the third low-$B$ magnetar.

The smallness of the period derivative is mirrored in the spin-inferred $B_{\rm dip}$ and $\tau_{\rm c}$, which is two or three orders of magnitudes larger than the typical values for magnetars ($10^3 - 10^5$\,yr). These two properties are suggestive that these three sources might be old magnetars which have already experienced a substantial field decay over their lifetime. Further features, such as the small number of detected bursts and the low quiescent luminosity, corroborate this interpretation. These discoveries showed how neutron stars with dipolar magnetic fields lower than those of ordinary magnetars can emit bursts and outbursts. Actually, one of the key ingredients to power magnetar bursting activity is the internal magnetic field, its toroidal component in particular.
Therefore, the low-$B$ magnetars should retain a strong-enough ($\sim10^{14}$\,G) internal toroidal field to seldomly produce crustal displacements, resulting in a much lower burst rate compared to younger objects. 
Indeed, magneto-thermal evolutionary models support this scenario: the evolution of an initial $B_{\rm dip} \sim 2 \times 10^{14}$\,G provides the observed characteristics for SGR\,0418+5729 and Swift\,J1822.3--1606 at an age of $\sim 1$\,Myr and 0.5\,Myr, respectively, if the initial toroidal field is high enough, $\sim 10^{16}$\,G for the former and $\sim 5 \times 10^{15}$\,G for the latter.

\begin{figure}
\centering
\includegraphics[width=0.95\columnwidth, trim = 1cm 5cm 1cm 0cm, clip]{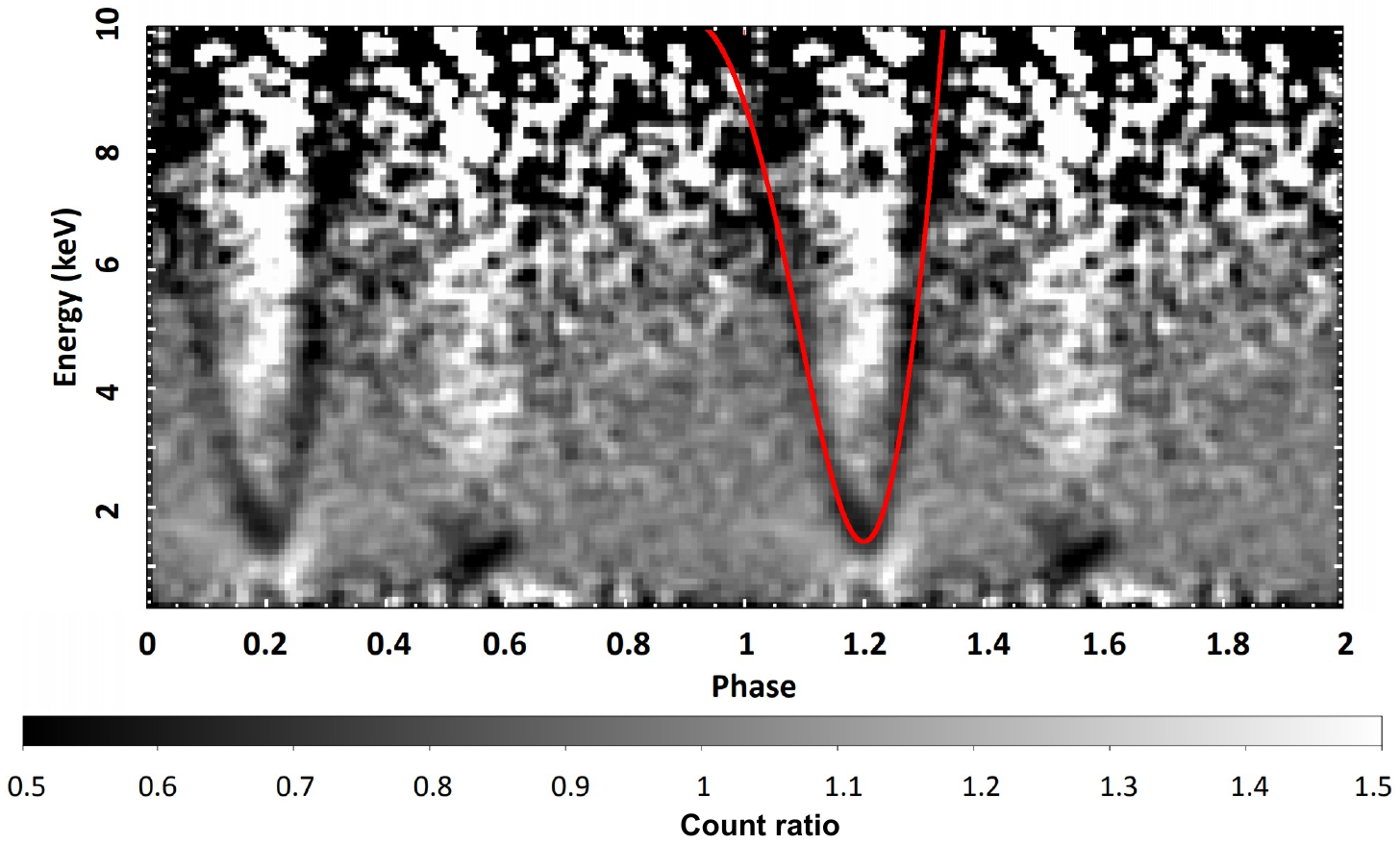} 
\caption{\label{fig:mag_feature} Normalised energy versus phase image of SGR\,0418+5729 extracted from the {\it XMM-Newton} observation carried out on 2009 August 12, about two months after the outburst onset. The image was obtained by binning the source counts into 100 phase bins and 100-eV-wide energy channels and then by normalising the counts to the phase-averaged spectrum and pulse profile. The red line indicates the expected phase dependence of the features in the proton cyclotron scattering model developped by \cite{Tiengo13}. From \cite{Tiengo13}.}
\end{figure}

An observational gauge of the magnetic field strength in neutron stars is provided by cyclotron features in their X-ray spectra. Such a feature was reported in the low-$B$ magnetar SGR\,0418+5729 during its 2009 outburst \cite{Tiengo13}. The most striking property of the absorption line is its strong dependence on the star rotational phase: 
the line energy is in the $1-5$\,keV range and varies strongly with the spin phase, roughly by a factor of 5 in one-tenth of a spin cycle (see Fig.\,\ref{fig:mag_feature}). The most plausible explanation for the feature variability is given if the line is due to cyclotron resonant scattering. The cyclotron energy for a particle of charge $e$ and mass $m$ is
\begin{equation}
    E_{\rm cycl} = \frac{11.6}{1+z} \left( \frac{m_{\rm e}}{m} \right) B_{\rm 12}\;\rm{keV},
\end{equation}
where $z \sim 0.8$ is the gravitational redshift, $m_{\rm e}$ is the electron mass, and $B_{\rm 12}$ is the magnetic field in units of $10^{12}$\,G. In this scenario, the strong dependence of the line energy with phase arises from the different fields experienced by the charged particles interacting with the photons emitted from the surface. If protons are the particles responsible for the scattering, the feature energy implies a magnetic field spanning from $2 \times 10^{14}$\,G to $> 10^{15}$\,G, much higher than $B_{\rm dip}$. According to the simplest model, the scattering may take place in a small localised magnetic loop close to the surface of the neutron star. If this interpretation is correct, this result provides evidence for a complex topology of the magnetosphere, with strong non-dipolar magnetic field components. A similar phase-dependent absorption feature was detected in another low-$B$ magnetar, Swift\,J1822.3--1606 \cite{Rodriguez16}. The energy of the feature varies between $\sim 5$ and $\sim 12$\,keV and the resulting strength of the magnetic field in the small magnetic coronal loop is in the $(6 - 25) \times 10^{14}$\,G range. Two additional detections were reported in two X-ray dim isolated neutron stars (see Sec.\,\ref{subsec:xdins_properties}).

\subsection{Magnetar-like activity from high-$B$ rotation-powered pulsars}
\label{subsec:highb_psr}

As there are magnetars having rather low dipolar magnetic fields, there exists a small group of rotation-powered pulsars with magnetar field strengths. Since the strong magnetic field is the source of instabilities in magnetars, magnetar-like bursts and outbursts might also be expected from \hbindex{high-$B$ pulsars} and, indeed, have been observed in two such sources. 

PSR\,J1846$-$0258 is the neutron star associated with the supernova remnant Kes\,75. With a rotation period of 326\,ms, it is one of the youngest known pulsars in the Galaxy and has a $B_{\rm dip} \sim 5 \times 10^{13}$\,G. PSR\,J1846$-$0258 behaved as a typical rotation-powered pulsar for the majority of its observed lifetime, with an X-ray luminosity lower than its spin down luminosity and powering a bright pulsar wind nebula. However, it has no detectable radio emission, although this may be due to unfavourable beaming. Data collected with the {\it Rossi X-Ray Timing Explorer} between 2000 and 2006 allowed us to measure the braking index equal to $n = 2.65 \pm 0.01$ \cite{Livingstone06}. This value is less than 3, which is expected from magnetic dipole radiation, implying that another physical mechanism must affect the pulsar rotational evolution (e.g., a time-varying magnetic moment or the effects of magnetospheric plasma on the spin-down torque). Remarkably, this pulsar emitted a few short ($\sim 0.1$\,s) magnetar-like bursts in 2006 May coincident with an enhancement of the pulsed flux that lasted about 2 months \cite{Gavriil08}. This event recalled a typical magnetar outburst. Its quiescent spectrum is like that of other young rotation-powered pulsars, well described by a simple power law with photon index $\Gamma \sim 1.2$. During the outburst, the spectrum softened ($\Gamma \sim 1.9$) significantly and called for an extra component, a blackbody with $kT \sim 0.9$\,keV, resembling the spectra of magnetars. Due to the softening, the $0.5-2$\,keV flux showed the largest increase (a factor of $\sim20$), whereas the $2-10$\,keV flux increased only by a factor of $\sim6$. The onset of the outburst was accompanied by a large glitch and an increase in the timing noise of the pulsar. The glitch recovery was unusual for a rotation-powered pulsar, but was reminiscent of timing behaviour observed from magnetars: the glitch decayed over $\sim130$\,days resulting in a net decrease of the pulse frequency. 
A 7-year post-outburst monitoring campaign revealed a change in the braking index. The post-outburst value, $n = 2.19 \pm 0.03$, is discrepant at the $14.5\sigma$ level from the pre-outburst braking index. A change in $n$ might be due to a change in the configuration of the pulsar magnetosphere \cite{Archibald15}. 
After 14 years of quiescence, PSR\,J1846$-$0258 entered again in outburst on August 1st, 2020 \cite[see e.g.,][]{Blumer21}. The spectral evolution was similar to what was observed previously in 2006. Radio magnetars are generally characterised by spin-down luminosity $\dot{E}$ larger than their X-ray luminosity $L_{\rm X}$, in line with the rotation-powered pulsars, and radio pulsations had been detected within a few days of the outburst onset. For PSR\,J1846$-$0258, the X-ray efficiency $L_{\rm X}/\dot{E} << 1$; therefore, if its radio emission is similar to those of the radio magnetars, the source should have become radio loud. However, radio observations performed within 5 days of the 2020 outburst onset did not detected any emission.

The only other radio pulsar where magnetar-like activity was observed so far is PSR\,J1119$-$6127, located at the center of the supernova remnant G292.2$-$0.5. The period $P \sim 408$\,ms and the spin-down rate $\dot{P} \sim 4 \times 10^{-12}$\,s\,s$^{-1}$ imply a characteristic age $\tau_{\rm c} \sim 1.9$\,kyr and dipolar magnetic field $B_{\rm dip} \sim 4 \times 10^{13}$\,G. In the radio band, it usually behaves as a stable radio pulsar, with the exception of an unusual glitch followed by a short-lived change in the pulse profile in 2007. In 2002, a {\it Chandra} observation unveiled the X-ray counterpart of the rotation-powered pulsar and a faint compact pulsar wind nebula around it. 
Follow-up studies resolved the point-like source from the diffuse emission. The combination of a power law ($\Gamma \sim 1.9$) and a blackbody ($kT \sim 0.2$\,keV) provided a satisfactory description of the pulsar spectrum \cite{Safiharb08}. X-ray pulsations were detected below 2.5\,keV with a single-peaked pulse profile, which is phase-aligned with the radio one. Pulsations were also detected at $\gamma$-rays, making it the $\gamma$-ray pulsar with the highest $B_{\rm dip}$. PSR\,J1119$–$6127 manifested itself as a magnetically active rotation-powered pulsar in 2016. On July 27th and 28th, 2016, the {\it Fermi} Gamma-ray Burst Monitor and {\it Swift} Burst Alert Telescope triggered on two short bursts from the direction of the pulsar. The field of view was observed about 70\,s after the second burst and a bright X-ray source was found at the position of PSR\,J1119$–$6127: a magnetar-like outburst started \cite{Archibald16}. At the outburst onset, a glitch took place and the pulsar turned off as a radio-loud source. After the reactivation in radio two weeks later, it displayed a magnetar-like radio spectral flattening and the radio profile changed to a two-peaked pulse profile. The established features of a magnetar outburst were witnessed: besides timing anomalies, a flux enhancement by a factor of $>160$ was registered, and the spectrum underwent a hardening with $kT$ increasing from $\sim0.2$\,keV to $\sim1.1$\,keV.

These two sources belong to a small yet growing class of objects that straddle the boundary between rotationally and magnetically powered neutron stars. In the coming years, additional high-$B$ pulsars might undergo a magentar-like transition, strengthening the link between radio pulsars and magnetars and, thus, providing observational evidence for a grand unification of the isolated neutron star population.

\section{Central Compact Objects}
\label{sec:cco}

\hbindex{Central Compact Objects} (CCOs) form a small, heterogeneous group of isolated X-ray emitting neutron stars, observed close to the geometrical center of young (0.3 -- 7\,kyr) supernova remnants and characterized by the absence of emission at other wavelengths (see \cite{Deluca17} for a review and references therein). This class consists of a dozen confirmed sources\footnote{See the complete catalog at \url{http://www.iasf-milano.inaf.it/~deluca/cco/main.htm}.} that only show thermal soft X-ray emission (0.2--5\,keV), with no evidence for a non-thermal component. Their spectra are well modelled by the sum of two blackbodies with temperatures $kT \sim 0.2-0.5$\,keV and small emitting radii ranging from 0.1 to a few km. Their luminosity, of the order of 10$^{33}$\,\lum, is generally steady, and in most cases pulsations are not detected (see Table\,\ref{tab:cco_list}).

The name CCO was used for the first time by Pavlov {\it et al.} \cite{Pavlov00} to refer to the point-like source at the centre of the supernova remnant Cassiopeia\,A. Since then, it had become the label for objects that share the above-mentioned properties, although it could be misleading. The Crab pulsar is located at the center of its nebula and is a compact object, however it is not classified as a CCO because of its emission at other wavelengths apart from X-rays. The first observed CCO, 1E\,161348--5055 in the supernova remnant RCW\,103, turned out to be a unique source, exhibiting a magnetar-like outburst in 2016 and possibly another one in 1999--2000 (see Sec.\,\ref{sec:rcw_outburst}).

\begin{table}[ht]
\begin{center}
\caption{Properties of the well-established central compact objects (above the line) and two candidates (below the line). Modified from \cite{
Gotthelf13}.}
\label{tab:cco_list}
\footnotesize{
\begin{tabular}{ccccccc}
\hline
CCO & SNR & Age$^a$ & $d$ & $P$  & $B_{\rm dip}$ & $L_{\rm x, bol}$  \\
    &     & (kyr) & (kpc) & (s) & ($10^{10}$\,G) & (\lum) \\
\hline
PSR\,J0821--4300        & Puppis\,A    & 4.5        & 2.2   & 0.112          & 2.9      & 5.6 $\times$ 10$^{33}$     \\
CXOU\,J085201.4--461753  & G266.1--1.2  & 1          & 1     & $\dots$        & $\dots$  & 2.5 $\times$ 10$^{32}$     \\
1E\,1207.4--5209         & G296.5+10.0  & 7          & 2.2   & 0.424          & 9.8      & 2.5 $\times$ 10$^{33}$     \\
CXOU\,J160103.1--513353  & G330.2+1.0   & $\leq$ 3   & 5     & $\dots$        & $\dots$  & 1.5 $\times$ 10$^{33}$     \\
1WGA\,J1713.4--3949      & G347.3--0.5  & 1.6        & 1.3   & $\dots$        & $\dots$  & $\sim$ 1 $\times$ 10$^{33}$ \\
XMMU\,J172054.5--372652  & G350.1--0.3  & 0.9        & 4.5   & $\dots$        & $\dots$  & 3.9 $\times$ 10$^{33}$    \\
XMMU\,J173203.3--344518  & G353.6--0.7  & $\sim$ 27  & 3.2   & $\dots$        & $\dots$  & 1.3 $\times$ 10$^{34}$ \\
PSR\,J1852+0040  & Kes\,79       & 7          & 7     & 0.105          & 3.1      & 5.3 $\times$ 10$^{33}$     \\
CXOU\,J232327.9+584842  & Cas\,A        & 0.33       & 3.4   & $\dots$        & $\dots$  & 4.7 $\times$ 10$^{33}$     \\
CXOU\,J181852.0--150213 & G15.9+0.2     & 1 -- 3     & 8.5   & $\dots$        & $\dots$  & $\sim$ 1 $\times$ 10$^{33}$ \\
1E\,161348--5055$^b$    & RCW\,103      & $\sim$ 2   & 3.3   & $24\times10^3$ & $\dots$  & 1.1$-$80  $\times$ 10$^{33}$ \\
\hline
2XMMi\,J115836.1--623516 & G296.8--0.3     & 10         & 9.6   & $\dots$  & $\dots$  & 1.1 $\times$ 10$^{33}$    \\
XMMU\,J173203.3--344518  & G353.6--0.7     & $\sim$ 27  & 3.2   & $\dots$  & $\dots$  & 1.3 $\times$ 10$^{34}$    \\
\hline
\hline
\end{tabular}
\begin{list}{}{}
\item[$^a$]{The age refers to the supernova remnant age.}
\item[$^b$]{1E\,161348--5055 has shown magnetar-like behaviour.}
\end{list}}
\end{center}
\end{table}

\subsection{Fun facts about CCOs}
\label{sec:cco_properties}

The proof that CCOs are indeed neutron stars came with the detection of pulsations from three objects: PSR\,J1852+0040 in Kes\,79 ($P \sim 105$\,ms; \cite{Gotthelf05}), PSR\,J0821--4300 in Puppis\,A ($P \sim 112$\,ms; \cite{Gotthelf09}), and 1E\,1207.4--5209 in G296.5+10.0 ($P \sim 424$\,ms; \cite{Zavlin00}). Observational campaigns spanning several years made it possible to measure the corresponding period derivatives of the order of $10^{-18}-10^{-17}$\,s\,s$^{-1}$ \cite[see e.g.,][]{Gotthelf13}. From the derived timing solutions, we can infer that {\it (1)} the spin down luminosity is about a factor of 10 lower than the X-ray luminosity; {\it (2)} the characteristic age $\tau_{\rm c}$ is $4-5$ orders of magnitude larger than the age of the host supernova remnants, indicating that CCOs were either born spinning at nearly their present periods or had an atypical magnetic field evolution; and {\it (3)} the inferred dipolar magnetic field $B_{\rm dip}$ is $10^{10}-10^{11}$\,G, remarkably smaller than that of the bulk of the radio pulsars. Given these low $B$-field strengths, CCOs were labelled as `anti-magnetars': neutron stars born with weak magnetic fields that have not been effectively amplified trough dynamo effects, due to their slow rotation at birth. The `anti-magnetars' scenario does not, however, account for several observational aspects.

\begin{figure}
\centering
\includegraphics[width=1.\columnwidth,trim={0cm 5cm 0cm 0cm},clip=true]{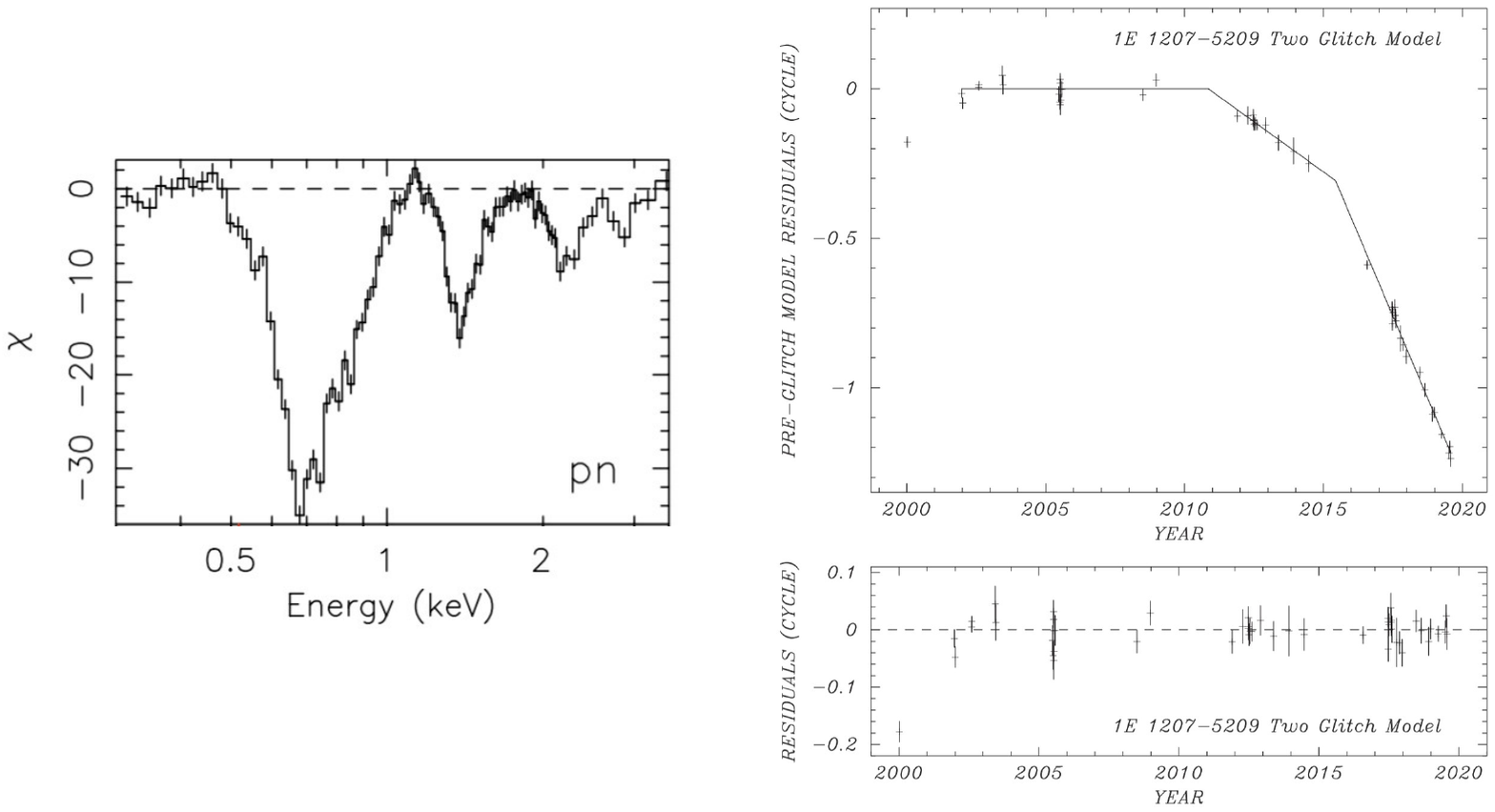}
\caption{{\it Left:} Residuals in units of sigma obtained from a comparison between the data with the best fit thermal continuum model (i.e., the sum of two blackbodies) for 1E\,1207.4--5209. The presence of four absorption spectral features at $\sim 0.7$\,keV, 1.4\,keV, 2.1\,keV, and 2.8\,keV is evident. From \cite{Deluca04}. {\it Right:} Phase residuals from the pre-glitch timing solution, modeled as two successive glitch-like changes in frequency. Glitch epochs of 2010 November 09 and 2014 May 23 are estimated from the intersection of the respective pre- and post-glitch fits. From \cite{Gotthelf20}.}
\label{fig:cco_1e1207}
\end{figure}

In the following, we focus on the properties of the three pulsating CCOs.
Model fitting of the X-ray spectrum of PSR\,J1852+0040 with two blackbodies yields small emitting radii ($R_{\rm 1}=1.9$\,km and $R_{\rm 2}=0.45$\,km, for components with temperatures of $kT_{\rm 1} = 0.30$\,keV and $kT_{\rm 2} = 0.52$\,keV, respectively; \cite{Halpern10}). The thermal emission is highly modulated with a pulsed fraction of $\sim 64$\% and is characterized by a single broad pulse, whose shape does not appear to be a function of energy. These properties also suggest that the emitting area is small. Such tiny, hot spots are at odds with the inferred magnetic field value since surface temperature anisotropies are associated with the effects of much stronger magnetic fields. For PSR\,J0821--4300, the spectrum is well described by two blackbodies with the addition of a spectral feature, which can be modelled by either an emission line at $\sim 0.75$\,keV or two absorption lines at $\sim 0.46$\,keV and $\sim 0.92$\,keV. The pulse profile of PSR\,J0821--4300 shows a sinusoidal shape with a 180$^{\circ}$ phase reversal at $\sim 1.2$\,keV (see e.g., \cite{Gotthelf13}). A pair of antipodal hot spots of different areas and temperatures on the neutron star surface is able to reproduce the thermal spectrum and the energy-dependent pulse profile. As for the case of the CCO in Kes\,79, explaining the non-uniform temperature distribution of the source may require a crustal field that is stronger than the external dipole field. Finally, the X-ray spectral energy distribution of 1E\,1207.4--5209 exhibits a continuum emission of thermal origin defined by the sum of two blackbodies. A satisfactory fit is achieved by including four absorption lines with simple Gaussian profiles centered at $\sim 0.7$\,keV, 1.4\,keV, 2.1\,keV, and 2.8\,keV (see Fig\,\ref{fig:cco_1e1207}, left panel; \cite{Bignami03, Deluca04}). The harmonically spaced spectral features are naturally explained by electron cyclotron absorption from the fundamental and three harmonics. This interpretation allows a direct measurement of the magnetic field: the fundamental cyclotron energy of $\sim 0.7$\,keV yields a magnetic field strength of $B_{\rm cyc} \sim 8 \times 10^{10}$\,G, assuming a 25\% gravitational redshift. This result is consistent with the magnetic field inferred from the spin down parameters, $B_{\rm dip} \sim 9.8 \times 10^{10}$\,G. The X-ray pulsation is dominated by the complicated modulation of the spectral features as a function of the rotational phase, while the continuum is almost unpulsed. Twenty years of timing observations performed between 2000 and 2019 revealed that 1E\,1207.4--5209 had two small glitches (see Fig.\,\ref{fig:cco_1e1207}, right panel; \citep{Gotthelf20}). This is the first time to detect such phenomenon in a CCO and in an isolated neutron star with a period derivative as small as that of this source. The phase ephemeris can be well modelled including two glitches that took place at the estimated epochs of November 9th, 2010, and May 23rd, 2014. Alternative timing models, which do not consider glitches, can also fit the data, however the resulting timing residuals and second frequency derivative are orders of magnitude larger than in isolated neutron stars with similar spin-down paramenters. No changes were observed in the spectrum and the central energies of the spectral features, meaning that the surface magnetic field strength was constant before and after the glitches. Therefore, it is interesting to consider that the glitches might be triggered by the motion of an internal field, as strong as those of canonical young pulsars that show glitches.\\

Beside the three pulsed CCOs, this group includes about ten more sources without detected pulsations. These objects have similar spectral properties as the CCO pulsars, making it reasonable to assume that they are also isolated neutron stars born with weak dipole magnetic fields. 
The most studied CCO is the X-ray point source at the center of the supernova remnant Cassiopeia\,A. Discovered in the first-light observations of {\it Chandra}, it is among the youngest-known neutron stars, as the supernova remnant age is estimated at $\sim$350\,yr. Its spectrum is well described by different models. A blackbody, a magnetic or non-magnetic hydrogen atmosphere models are consistent with the emission coming from small hot spots, similar to what has been observed for other CCOs. However, timing investigations were unsuccessful in identifying pulsations down to a pulsed fraction limit of 12\%. These apparently contradictory results are reconciled by the discovery that a low magnetized ($B_{\rm dip}<10^{11}$\,G) carbon atmosphere model gives a satisfactory spectral fit with the emission arising from the entire neutron star surface, which would not necessarily vary with the star rotation \cite{Heinke10}. A decrease of the surface temperature by 4\% and a 21\% change of the flux were reported over a time span of 10 years (from 2000 to 2009) by using multi-epoch {\it Chandra} observations. Such direct measurement of the cooling of an isolated neutron star would have profound theoretical implications for our understanding of the internal composition and structure of these compact objects. By applying cooling models, it would be possible to constrain neutrino emission mechanisms and envelope compositions \cite{Heinke10}. 
However, it was noted that these {\it Chandra} observations suffered from several instrumental effects that can cause time-dependent spectral distortions. Therefore, additional {\it Chandra} poitings were carried out in such a set-up to minimize the instrument effects in four epochs (2006, 2013, 2015, and 2020; \cite{Posselt22}). An apparent increase in the cooling rate between 2015 and 2020, and the variations of the inferred hydrogen column densities between the different epochs were reported. The authors note that these changes could indicate systematic effects such as caused by, for instance, an imperfect calibration of the increasing contamination of the optical-blocking filter.

The carbon atmosphere model with low magnetic field provides a good spectral fit not only for the CCO in Cassiopeia\,A, but also for a few more sources (see e.g., \cite[][]{Klochkov15}). These models assume homogeneous temperature distribution on the entire neutron star surface, naturally explaining the absence of detected pulsations within the current limit. Moreover, they make it possible to reconcile thermal luminosities with the known or estimated distances, and to constrain the neutron star equation of state. Mass and radius are used to estimate the surface gravity and gravitational redshift affecting the atmosphere model, and they are two of the four free fit parameters of the model. For instance, for the CCO in the supernova remnant G353.6--0.7 (as well known as HESS\,J1731--347), the obtained best-fit neutron star mass and radius are 1.55\,$M_{\odot}$ and 12.4\,km for the preferred distance of 3.2\,kpc. These values are compatible with the most commonly used nuclear equations of state \cite{Klochkov15}. Combining these results with cooling theory has a potential to put more stringent constraints (see for more details, \cite[][]{Ofengeim15}). \\

Observational results, such as the presence of hot spots in the pulsed CCOs and the occurrence of a glitch, hint at a stronger magnetic field in the interior of CCOs. Alternative explanations to the `anti-magnetar' hypothesis have been discussed. CCOs might be magnetars in quiescence characterised by a weak dipole field and a strong crustal magnetic field that emerges locally in confined areas. However, this scenario seems unlikely owing to the lack of variability in CCOs, at variance with what seen in magnetars, even if it could be correct for a sub-group of the sample. Another theory regarding the CCO nature posits that they are born with a canonical neutron star magnetic field that was buried by the fallback of the debris of the supernova explosion, the so-called `hidden magnetic field' interpretation. Several models proposed that a typical magnetic field in the range of $10^{12}-10^{14}$\,G can be pushed deep inside the crust by accreting a mass of $\sim 10^{-4}-10^{-2}$\,M$_\odot$. The result is an external magnetic field lower than the internal hidden one, which might re-emerge on a time scale of $10^{3}-10^{5}$\,yr once the accretion stops. After this stage, the magnetic field at the surface is restored close to its value at birth (see e.g., \cite{Torresforne16}). The thermal evolution during the re-emergence phase can produce different degrees of anisotropy in the surface temperature, explaining the large pulsed fraction measured for the CCO in Kes\,79 and the antipodal hot spots of PSR\,J0821--4300 in Puppis\,A. Furthermore, these models predict that weak dipolar magnetic fields could be common in very young ($<$ few kyr) neutron stars, and CCOs might be ancestors of old radio pulsars as long as their surface field grows to the critical limit required for radio emission. An observational test to prove the `hidden magnetic field' scenario is to measure the braking index $n$. In case of a constant magnetic field and magneto dipolar braking mechanism, $n$ is equal to 3; while values smaller than 3 translate into a growing field. Estimates of $n$ have not been achieved so far for CCOs; their low period derivatives make this task extremely challenging. Interestingly, all reported values for young rotation-powered pulsar ($\tau_{\rm c} \sim 10^3-10^4$\,yr) are in agreement with $n<3$, providing support for this picture (see e.g., \cite{marshall16}).

\subsection{1E\,161348--5055: a hidden magnetar}
\label{sec:rcw_outburst}

1E\,161348$-$5055 (1E\,1613) was discovered in 1980 when the {\it Einstein} Observatory detected a point-like source lying close to the center of the 2-kyr-old supernova remnant RCW\,103. It was the first candidate radio-quiet isolated neutron star discovered in a supernova remnant. It has been one of the prototypes of the CCO class. However, in the last two decades, observations revealed remarkable features that make 1E\,1613 stand out from the other CCOs. Firstly, unlike the other members of this class that generally have a steady emission, it displays a strong flux variability on a month/year time scale, undergoing an outburst at the end of 1999 with a flux increase by a factor of $\sim 100$. Secondly, a long periodicity of 6.7\,hr was detected with a strong nearly-sinusoidal modulation \cite{Deluca06}. Although the 6.7-hr periodicity could be recognized in all the data sets that were long enough, the corresponding pulse profile changed according to the source flux level: from a sine-like shape when the source is in a low state (observed soft X-ray flux $\sim 10^{-12}$\,\flux), to a more complex, multi-peak configuration in a high state ($\sim 10^{-11}$\,\flux). The long-term variability, the unusual periodicity, and the pulse profile changes have contributed to build an intriguing phenomenology in the neutron star scenario. Based on these characteristics, two main interpretations were put forward: 1E\,1613 could be either the first low-mass X-ray binary in a supernova remnant with an orbital period of 6.7\,hr or a peculiar isolated compact object with a rotational spin period of 6.7\,hr, invoking the magnetar scenario that naturally accounts for the flux and pulse shape variations.

\begin{figure}
\centering
\includegraphics[width=1.\columnwidth,trim={16cm 6cm 0cm 0cm},clip=true]{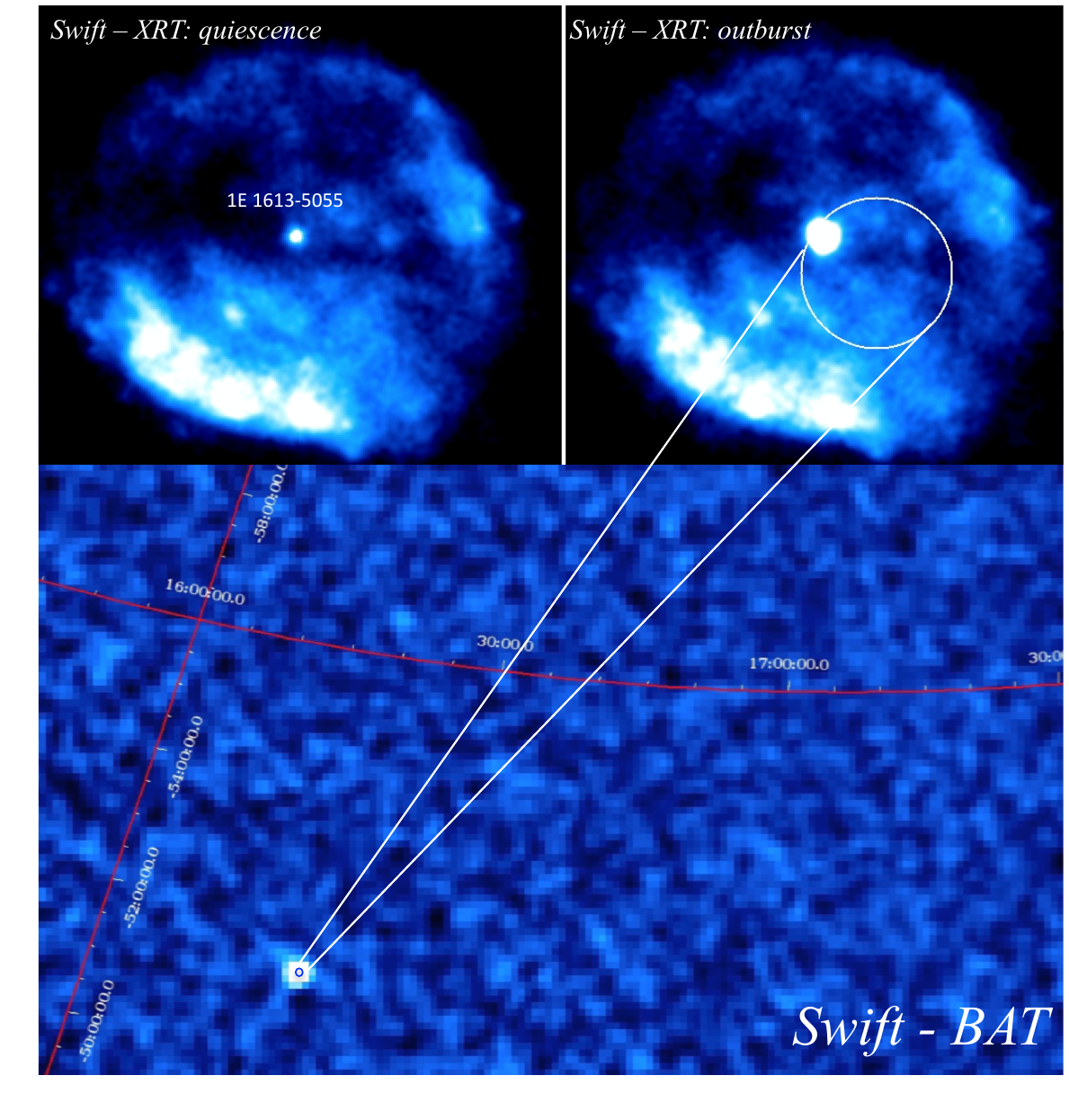}
\caption{{\it Swift}-BAT 15 -- 150\,keV image of the burst detected on 2016 June 22 from the direction of the supernova remnant RCW\,103 (bottom). Two {\it Swift}-XRT co-added 1 -- 10\,keV images of RCW\,103 during the quiescent state of 1E\,161348$-$5055 (from 2011 April 18 to 2016 May 16; exposure time of 66\,ks; top left) and in outburst (from 2016 June 22 to July 20; exposure time of 67\,ks; top right). The white circle marks the positional accuracy of the detected burst, which has a radius of 1.5 arcmin. From \cite{Rea16}.}
\label{fig:cco_outburst}
\end{figure}

In 2016, a new event shed light on the nature of this source: on June 22nd, the {\it Swift} Burst Alert Telescope triggered on a short ($\sim 10$\,ms) hard X-ray magnetar-like burst from the direction of RCW\,103 (see e.g., \citep{Rea16}), with a spectrum described by a blackbody ($kT \sim 9$\,keV) and a luminosity of $\sim 2 \times 10^{39}$\,\lum\ in the 15 -- 150\,keV energy range for a distance of 3.3\,kpc. Meanwhile, the {\it Swift} X-Ray Telescope detected an enhancement of the 0.5 -- 10\,keV flux by a factor of $\sim 100$ with respect to the quiescent level that persisted for years and was observed up to one month before (see Fig.\,\ref{fig:cco_outburst}). Follow-up observations with {\it Chandra} and {\it NuSTAR} confirmed that the burst marked the onset of a magnetar-like outburst. For the first time, a hard X-ray, non-thermal spectral component was detected up to $\sim 30$\,keV, modelled by a power law with photon index $\Gamma \sim 1.2$, while the soft X-ray spectrum was well described by the sum of two blackbodies. The light curve displayed two broad peaks per cycle, in contrast with the sinusoidal shape observed since 2005. In the first year from the onset of the outburst, the overall energy emitted was $\sim 2 \times 10^{42}$\,erg \cite{Borghese18}. This event prompted searches for an infrared counterpart. The {\it Hubble Space Telescope} and the {\it Very Large Telescope} observed the field of view few weeks after the onset (see e.g., \cite{Esposito19}). The images disclose a new object at the position of the source, not detected in previous observations. The counterpart properties ruled out the binary scenario, however it is not clear whether the infrared emission comes from the neutron star magnetosphere or a fall-back disk. 

While all the aspects caught by these observations - the appearance of a hard power-law tail at the outburst peak, the variability of the pulse profile in time, and the infrared counterpart – point towards a magnetar interpretation of 1E\,1613, its long periodicity is puzzling; a very efficient braking mechanism is required to slow down the source from a fast birth period ($< 0.5$\,s) at the currently measured value of 6.7\,hr in $\sim 2$\,kyr. Most models consider a propeller interaction with a fall-back disk that can provide an additional spin-down torque besides that due to dipole radiation. Ho \& Andersson \cite{Ho17} predict a remnant disk of $\sim 10^{-9}$\,M$_\odot$ around a rapidly rotating neutron star that is initially in an ejector phase, and after hundreds of years its rotation period slows down enough to allow the onset of a propeller phase. By matching its observed spin period and young age, 1E\,1613 is found to have a slightly higher dipolar magnetic field than all known magnetars, $B_{\rm dip} \sim 5 \times 10^{15}$\,G.

\section{X-ray Dim Isolated Neutron Stars}
\label{sec:xdins}

Thanks to its high sensitivity in the soft X-ray band ($0.1-2.5$\,keV), the {\it ROSAT} satellite led to the discovery of seven thermally-emitting soft X-ray pulsars, labelled the \hbindex{X-ray Dim Isolated Neutron Stars} (XDINSs), and commonly nicknamed `The Magnificent Seven' (see \cite{Turolla09} for a review and references therein). Radio emission from these sources has not been detected so far despite deep searches, while all of them have confirmed optical and/or ultraviolet counterparts. XDINSs are hotter than they ought to be considering their ages and the available energy reservoir provided by the loss of rotational energy. They are among the closest neutron stars we know of, with distances $\leq500$\,pc as derived from the modelling of the distribution of the hydrogen column density $N_{\rm H}$. Parallax measurements are available for only two sources, RX\,J1856.5--3754 and RX\,J0720.4--3125, yielding a distance of $123^{+11}_{-15}$\,pc and $280^{+210}_{-85}$\,pc, respectively, in agreement with estimations obtained from the spectrum and $N_{\rm H}$. Unlike CCOs, XDINSs have strong  magnetic field ($\sim10^{13}$\,G), similar to those of the high-$B$ pulsars, are old objects ($\tau_{\rm c} \sim$ a few \,Myr), and are not found at the center of supernova remnants. Table\,\ref{tab:xdins_list} reports the overall properties of this class of isolated neutron stars.

\begin{table}[ht]
\begin{center}
\caption{Overall properties of the X-ray dim isolated neutron stars. $E_0$ refers to the central energies of the broad absorption line. $B_{\rm dip}$ corresponds to the surface, dipolar strength of the magnetic field measured at the equator and $B_{\rm cyc}$ is the magnetic field strength evaluated assuming that the spectral features in the phase-averaged spectra are proton cyclotron resonances. $L$ indicates the bolometric luminosity radiated from the surface. Adapted from \cite{Pires14}.}
\label{tab:xdins_list}

\footnotesize{
\begin{tabular}{@{}lcccccccc}
\hline
\hline
Source & \kt\  & $P$ & $\log{\dot{P}}$ & $\tau_{\rm c}$ & $E_0$ & $B_{\rm dip}$ 	&  $B_{\rm cyc}$ & $\log{L}$   \\ 
	   & (eV)  & (s) & 	   (\ss)      & 	     (10$^6$~yr)       & (eV)  & (10$^{13}$~G) & (10$^{13}$~G) & (\lum) \\
\hline
RX\,J1856.5--3754 & 61     & 7.06 & -13.527  & 3.8 & --  & 1.47 & -- & 31.5-31.7   \\
RX\,J0720.4--3125$^a$ & 84--94 & 8.39 & -13.156  & 1.9 & 311 & 2.45 & 5.62 & 32.2-32.4 \\
RX\,J1605.3+3249$^b$  & 100    & $\dots$ & $\dots$  & $\dots$ & 400 & $\dots$ & 8.32 & 31.9-32.2 \\
RX\,J1308.6+2127  & 100    & 10.31 & -12.951 & 1.4 & 390 & 3.48 & 3.98 & 32.1-32.2 \\
RX\,J2143.0+0654  & 104    & 9.43 & -13.398  & 3.7 & 750 & 1.95 & 1.41 & 31.8-31.9 \\
RX\,J0806.4--4123 & 95     & 11.37 & -13.260 & 3.2 & 486 & 2.51 & 9.12 & 31.2-31.4 \\
RX\,J0420.0--5022$^c$ & 48     & 3.45 & -13.553  & 1.9 & $\dots$ & 1.00 & $\dots$ & 30.9-31.0 \\  
\hline
\hline
\end{tabular}}
\end{center}
$^a$ Hambaryan {\it et al.} \cite{Hambaryan17} claimed that the most-likely genuine period is  16.78\,s, twice that reported in the literature.\\
$^b$ A period of 3.39\,s was claimed but not confirmed by later observations.\\
$^c$ An absorption line at $\sim0.3$\,keV was reported, but not confirmed.
\end{table}

\subsection{Overview of the observational properties}
\label{subsec:xdins_properties}

Timing studies of the XDINSs disclosed X-ray pulsations at spin periods in the range of $3-12$\,s with period derivatives of the order of $10^{-14}-10^{-13}$\,s\,s$^{-1}$, implying dipolar magnetic fields $B_{\rm dip} \sim (1-4) \times 10^{13}$\,G and characteristic ages $\tau_{\rm c} \sim 1-4$\,Myr. These properties place the XDINSs at the end of the rotation-powered radio pulsar towards long periods and below the magnetars in the $P-\dot{P}$ diagram (Fig.\,\ref{fig:p_pdot}). RX\,J1605.3+3249 is the only member of this neutron star group that still lacks a coherent timing solution. A possible candidate spin period of 3.39\,s and spin down derivative of $\sim 1.6 \times 10^{-12}$\,s\,s$^{-1}$ were proposed with a significance level of $\sim4\sigma$, making RX\,J1605.3+3249 the XDINS with the highest magnetic field ($B_{\rm dip} \sim 7.5 \times 10^{12}$\,G, \cite{Pires14}). Targeted monitoring campaigns with {\it XMM-Newton} and {\it NICER} ruled out this candidate with stringent upper limits on the pulsed fraction equal to 1.5\% and 2.6\% for periods above 150\,ms and 2\,ms, respectively (see e.g., \cite{Malacaria19}). Moreover, for RX\,J0720.4--3125 Hambaryan {\it et al.} \cite{Hambaryan17} claimed that the most-likely genuine period is twice that reported in the literature, 16.78\,s instead of 8.39\,s. A second peak was identified in the periodogram of all pointed {\it XMM-Newton} observations in different energy bands and, for some energy intervals, the timing series showed a more significant peak corresponding to $P=$16.78\,s. The light curves folded at the new claimed period display a markedly double-peaked shape that depends on time and energy. \\

\begin{figure}
\centering
\includegraphics[width=1.\columnwidth]{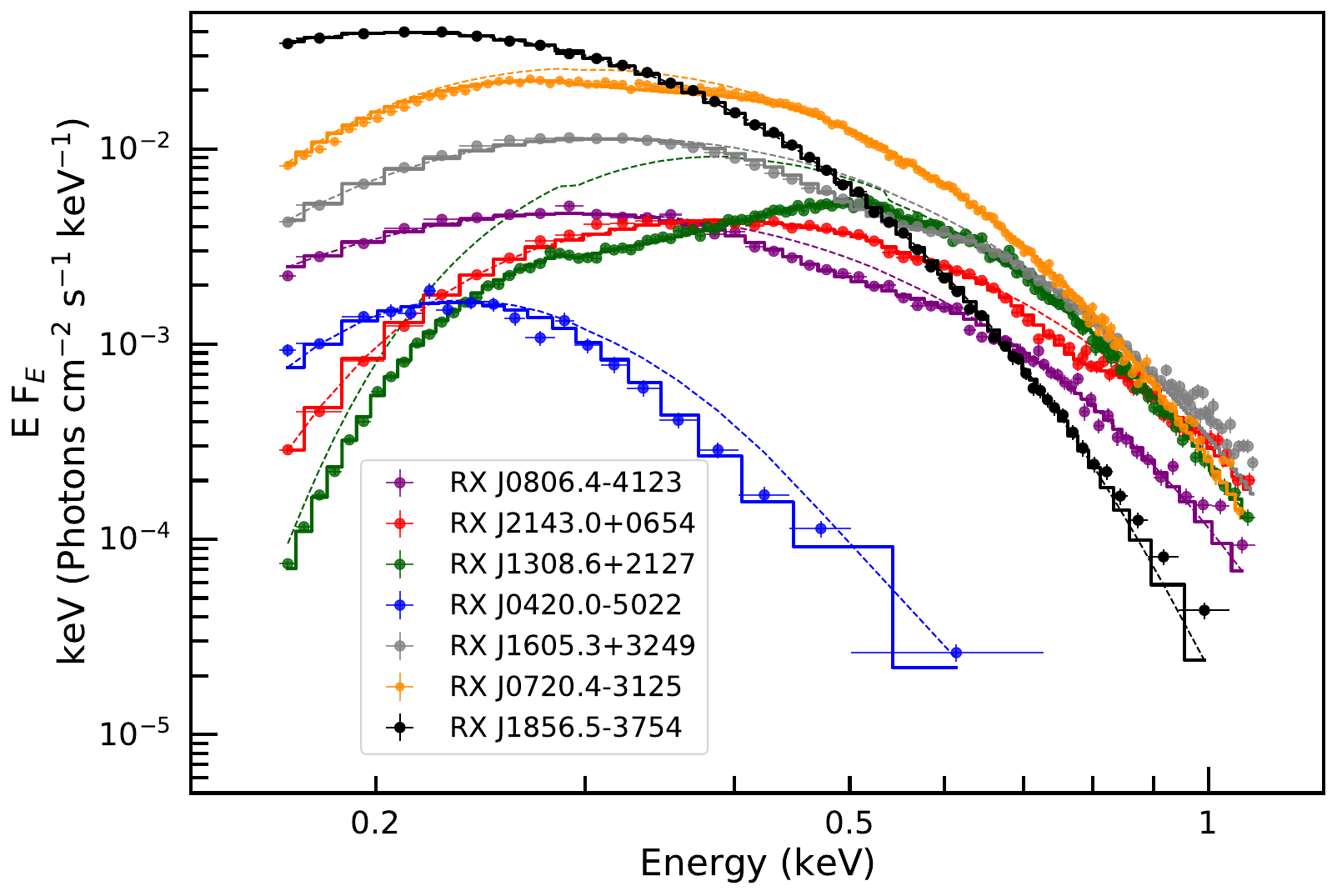}
\caption{Phase-averaged spectra of the X-ray dim isolated neutron stars extracted from the highest-counting statistic {\it XMM-Newton} observations. The solid lines represent the best-fitting models, consisting of an absorbed blackbody for RX\,J1856.5--3754 and an absorbed blackbody with the inclusion of a broad Gaussian line in absorption for all the other sources. The dashed lines indicate the blackbody components. Image credit: A. Borghese}
\label{fig:xdin_spec}
\end{figure}

XDINSs are characterized by very soft X-ray spectra described by an absorbed blackbody with low values for the absorption column density ($N_{\rm H} \sim 10^{20}$\,cm$^{-2}$) and temperatures in the range $50-100$\,eV, with no evidence for a power-law tail extending at higher energies (see Fig.\,\ref{fig:xdin_spec} and text below). The emission is purely thermal, with little contamination from magnetospheric activity or surrounding supernova remnant, and is believed to come directly from the neutron star surface. This trait makes `The Magnificent Seven' ideal targets for probing atmosphere models, and they can be used to constrain the mass-to-radius ratio neutron stars. For instance, fitting the phase-resolved spectra of RX\,J1308.6+2127 with a model consisting of a condensed iron surface and a partially ionized hydrogen-thin atmosphere above gives a true radius of $16 \pm 1$\,km for a standard neutron star mass 1.4\,M$_{\odot}$. The determined mass-to-radius ratio ($M$/M$_{\odot}$)/($R$/km) = $0.087\pm0.004$ implies a very stiff equation of state for this source \cite{Hambaryan11}. Applying the same procedure and model, the mass-to-radius ratio was derived for another XDINS, RX\,J0720.4--3125. Its ($M$/M$_{\odot}$)/($R$/km) is equal to $0.105\pm0.002$, assuming a mass of 1.4\,M$_{\odot}$ and radius of $13.3\pm0.5$\,km, inferred from spectral fitting. As in the previous measurement, this value points to a stiff equation of state \cite{Hambaryan17}. 

All the XDINSs have detected weak optical and ultraviolet counterparts thanks to {\it Hubble Space Telescope} observations \cite{Kaplan11}. The extrapolation at lower energies of the best-fit model inferred from the X-rays underestimates the actual observed flux. This phenomenon is known as optical excess, and is witnessed in all seven objects. Its origin is still an open issue. If the X-ray and optical radiation arise from different regions on the neutron star surface, variations in the excess are expected to correlate with changes in the X-ray pulsed fraction. However, such correlations have not been observed, suggesting that this picture is incomplete. Other alternative scenarios should be considered: the optical excess might be due to magnetospheric emission, atmospheric effects, or the presence of a nebula. Recently, an extended near-infrared emission was discovered around RX\,J0806.4--4123, having flux in excess with respect to the expected value from the extrapolation of its ultraviolet-optical flux. It can be interpreted as coming from a pulsar wind nebula fed by shocked pulsar wind particles of relatively low energy, or a disk with a favorable viewing geometry \cite{Posselt18b}. The spectral signature of the two scenarios is different and can be investigated with future high-resolution observations with the {\it James Webb Space Telescope}. Moreover, deep near-infrared surveys are required to understand whether such extended emission is a common property among some types of isolated neutron stars, or if RX\,J0806.4--4123 is a special case.\\

XDINSs were considered to be steady sources, hence RX\,J0720.4--3125 was included in the calibration targets for the EPIC and RGS instruments on board {\it XMM-Newton}. However, the monitoring campaign revealed long-term variations in its timing and spectral parameters. In the period between 2001 and 2003, while the total flux stayed constant, the blackbody temperature increased from $\sim84$\,eV to $\sim94$\,eV and the pulse profile changed as well, with an increase of the pulsed fraction. In the following years, this trend reversed with a decrease of the surface temperature, hinting at a cyclic behaviour. The temperature evolution covering the time span between 2000 and 2005 could be described by a sinusoidal function with a period of $\sim7$\,yr. The most likely explanation involved a freely precessing neutron star with two hot spots with different temperature and size, not located in antipodal positions. The observed blackbody temperature variations are produced by the changes in the viewing geometry of the two spots as the star precesses \cite{Haberl06}. RX\,J0720.4--3125 continued to be monitored with {\it XMM-Newton} and {\it Chandra} until 2012. The new observations from 2005 to 2012 disclosed a monotonic decrease in the temperature, excluding a cyclic pattern with a 7-yr period. Moreover, a 14-yr period seemed unlikely because the extrapolated temperature value at the end of the cycle was different from the initial one. On the other hand, the timing behaviour favoured the interpretation in terms of a single sudden event, such as a glitch, with the spectral parameters changing remarkably around the proposed glitch date \cite{Hohle12}. RX\,J0720.4--3125 is the only XDINS to have undergone a glitch so far.\\ 

As stated above, the hallmark of the XDINSs is their thermal spectrum detected in the soft X-rays without any need to include a power law at higher energies. Although a blackbody model provides an overall good description of their spectral energy distribution, broad absorption features had been found in most of the XDINSs (see Fig.\,\ref{fig:xdin_spec} and Table\,\ref{tab:xdins_list}). The brightest XDINS, RX\,J1856.5--3754, exhibits a featureless spectrum, compatible with a blackbody with hints of a second colder thermal component detected also in optical and ultraviolet bands \cite{Sartore12}. The presence of a spectral line at $\sim0.3$\,keV was initially claimed but not confirmed by later, deeper observations for the faintest member, RX\,J0420.0--5022 \cite{Kaplan11a}. For the remaining sources, the properties of the spectral features are similar: the central energies range from $\sim300$\,eV to $\sim800$\,eV, they are quite broad with widths generally in the range of $\sim70-170$\,eV, the equivalent widths are several tens of eV ($30-150$\,eV), and they vary with the spin phase. Note that the two XDINSs with no apparent spectral features are the coldest, with $kT\sim45-60$\,eV, while the other five objects with claimed lines have similar temperatures ($kT\sim80-100$\,eV). Deviations from a pure blackbody can be explained by several physical mechanisms. One hypothesis is that the spectral lines can be produced by proton cyclotron resonances in a hot ionized layer near the surface. The line energies imply a magnetic field strengths of the order of $10^{13}$\,G, in rough agreement with the values inferred from the timing. An alternative hypothesis would rest upon atomic transitions in a magnetised atmosphere. The energies of the absorption features can be matched with transitions in a hydrogen atmosphere in most cases, apart from one: for RX\,J2143.0+0654, the line energy of $\sim0.7$\,keV is substantially higher than what is observed for all the other sources, and for any transition in hydrogen, therefore a helium or heavier element atmosphere is required. Finally, an inhomogeneous surface temperature distribution can induce spectral distortions in the form of spectral features \cite{Vigano14}. Temperature anisotropies can be inferred from the measurements of small blackbody emitting areas and are theoretically expected, e.g. by magnetospheric particle bombardments or anisotropic thermal conductivity caused by a strong magnetic field. Predicting the temperature distribution on the neutron star surface is hard due to many theoretical uncertainties, such as the magnetic field geometry. Therefore, the easiest approach is to explore a wide variety of temperature profiles that differ in crustal temperature, magnetic field strength, and number of hotspots, and to compare them with the observations. Vigan\`o {\it et al.} \cite{Vigano14} concluded that two requirements are needed to produce a spectral line: one or more hot, small regions should have a temperature larger by at least a factor of $\sim2$ than the average temperature of the rest of the neutron star surface, and the hot spot(s) and the colder part should contribute equally to the total flux. Temperature inhomogeneities can play an important role in the emission processes and should be taken into account in combination with more sophisticated emission models (e.g., atmospheric/condensed surface models).

The discovery of a phase-dependent absorption feature in two low-magnetic field magnetars (see Sec.\,\ref{subsec:lowb_mag}), and the fact that XDINSs are believed to be descendants of magnetars, according to magneto-thermal evolutionary models \cite{Vigano13}, motivated the search for such features in `The Magnificent Seven'. Only in two sources, RX\,J0720.4--3125 and RX\,J1308.6+2127, was a phase-variable absorption feature found with properties similar to those of the line reported in the magnetars SGR\,0418+5729 and Swift\,J1822.3--1606, strengthening the evolutionary link between these two groups of isolated neutron stars \cite{Borghese15, Borghese17}. In both detections, the line energy is about $\sim0.75$\,keV with an equivalent width of $\sim15-30$\,eV. The features are significantly detected in only 20\% of the star rotational phase and appear to be stable over the time interval covered by the observations ($\sim12$\,yr and 7\,yr for RX\,J0720.4--3125 and RX\,J1308.6+2127, respectively). In light of the similarities with SGR\,0418+5729, the feature might be explained invoking by the same physical mechanism: proton cyclotron resonant scattering. In this scenario, the sharp variation with phase is ascribed to the presence of small-scale ($\sim100$\,m) magnetic structures close to the neutron star surface. The implied magnetic field in the loop is about a factor of $\sim5$ higher than the dipolar component for both sources. These findings are supportive of a picture in which the magnetic field of highly magnetised neutron stars is more complex than a pure dipole with deviations on a small scale, such as localized high $B$-field bundles. 

\section{Rotating Radio Transients}
To conclude the gallery tour of the non-accreting neutron stars, we briefly discuss the Rotating Radio Transients (RRATs), which were discovered only a few years ago through the detection of sporadic single radio pulses. The largest common denominator between the arrival times of these pulses made it possible to figure out an underlying periodicity pointing to neutron stars for these sources. The known population of RRATs tallies to about one hundred sources, and for one-fifth of them, it has been possible to also estimate the spin down rate. Even though there is evidence of RRATs with longer periods, older ages and higher magnetic fields than the average in RPPs, they are not tightly clustered or located in unusual areas in the $P$--$\dot{P}$ diagram. The current consensus, is that RRATs are radio-emitting RPPs that for some reason display an extreme form of the more common nulling observed in several radio pulsars. 

Only one RRAT, J1819--1458 ($P=4.3$\,s, $\tau_\mathrm{c}=0.1$\,Myr, $B\sim5\times10^{13}$\,G), has been observed as an X-ray pulsar. Its X-ray spectrum is thermal, with a blackbody temperature of 140\,eV, similar to that of some middle-age pulsars and of the XDINs, and a possible absorption feature at 0.5 keV. Interestingly, the distance to the source inferred from the radio dispersion measure implies an X-ray luminosity of $\approx4\times10^{33}$\,\lum, ten times larger than the spin-down power.

\section{Final remarks}
\label{sec:conclusion}

Although we have been observing neutron stars and pulsars at all wavelengths for more than fifty years, they are still delivering striking surprises, such as the recent discovery of powerful millisecond-long radio bursts from the magnetar SGR\,J1935+2154. This is largely due to the arrival of ever more powerful ground-based and space-borne observatories, but it is fair to say that the neutron star phenomenology exceeded the expectations of the astronomers. Indeed, at first glance, they appear to be simple objects: they are gravitationally collapsed bodies made predominantly of neutrons (and chances are that they are all ruled by the same--yet elusive---equation of state), nearly collapsing into a black hole, being just a few times their Schwarzschild radius. 
More than thirty years after Baade and Zwicky \cite{Baade1934} suggested that a supernova is the transition of an ordinary star to a neutron star, was it realized that neutron stars might have been observable for attributes other than the tremendous heat with which they emerge from the explosion \cite{Tsuruta1966}. 
 
Pacini \cite{Pacini1967} was probably the first to note that if neutron stars were endowed with strong magnetic fields (as could be expected from some considerations on the retainment of the magnetic moment of the progenitor star, but see, e.g., \cite{Gourgouliatos2018} and references therein), then they should have been powerful emitters of electromagnetic waves owing to their fast rotation (eq.\,\ref{eq:Edotdip}).

However, there is increasing evidence that the magnetic field is not simply a mean for pulsars to convert their rotational kinetic energy in radiation and particles. Magnetars in particular have been of paramount importance in suggesting that it is the magnetic field that drives the amazing observational diversity of the different classes of neutron stars, not only the external poloidal component, which is responsible for the rotation braking, but also the toroidal internal component (which may store much more energy), their balance, and their (co)evolution. The existence of this component in any kind of neutron star is necessary because purely poloidal configurations are intrinsically unstable on short scales (as well as the purely toroidal ones), as is suggested by the model of dynamo formation of the magnetic field in the proto-neutron star, and it is tempting to invoke it to answer, for instance, the unusual behaviour of some pulsars or the plentiful puzzling radio emission of the recycled millisecond pulsars. Furthermore, observational evidence of surface magnetic fields more complex than dipoles is emerging. This hidden factor in the $P$--$\dot{P}$ diagram may be the \emph{fil rouge} connecting the various groups of neutron stars and explaining their diversity together with a few other fundamental characteristics, such as the age, the mass, and the envelope composition.  

Huge theoretical efforts are ongoing on neutron stars. Moreover, upcoming new instruments will help define their equation of state, characterize their emission properties, unveil the existence of new mysterious objects, and finally, constrain the census of slippery objects, such as the magnetars, the RRATs, and transient radio pulsars in general, and thus, the Galactic and nearby extragalactic population. However, another awesome aspects of pulsars is that even if we still have to figure out a lot about them, we can nonetheless use them to investigate many problems of physics. As they are clocks with extreme properties in extreme environments, they can be used to test general relativity and alternative theories (e.g. \cite{Stairs2003}), to probe the interstellar medium and magnetic field, and to explore the low-frequency gravitational background. Though we do not understand them well, pulsars are going to bring us gifts and keep us busy for many more decades to come!

\newpage

\bibliographystyle{ieeetr} 
\bibliography{biblio.bib}

\end{document}